\documentclass[utf8]{FrontiersinHarvard} 

\newcommand{\ilam}{erg cm$^{-2}$ s$^{-1}$ sr$^{-1}$ \AA$^{-1}$}
\newcommand\arcsec{\mbox{$^{\prime\prime}$}}%
\usepackage{url,hyperref,lineno,microtype,subcaption}
\usepackage[onehalfspacing]{setspace}
\usepackage{lscape}

\def\keyFont{\fontsize{8}{11}\helveticabold }
\def\firstAuthorLast{Kowalski} 
\def\Authors{Adam F. Kowalski\,$^{1,2,3,*}$}
\def\Address{$^{1}$National Solar Observatory, University of Colorado Boulder, 3665 Discovery Drive, Boulder, CO 80303, USA;
$^{2}$Department of Astrophysical and Planetary Sciences, University of Colorado, Boulder, 2000 Colorado Ave, CO 80305, USA;
$^{3}$Laboratory for Atmospheric and Space Physics, University of Colorado Boulder, 3665 Discovery Drive, Boulder, CO 80303, USA.}
\def\corrAuthor{Adam F. Kowalski}

\def\corrEmail{adam.f.kowalski@colorado.edu}

\begin{document}
\onecolumn
\firstpage{1}

\title[NUV Flare Models]{Near-Ultraviolet Continuum Modeling of the 1985 April 12 Great Flare of AD Leo}

\author[\firstAuthorLast ]{\Authors} 
\address{} 
\correspondance{} 

\extraAuth{}

\maketitle

\begin{abstract}
White-light stellar flares are now reported by the thousands in long-baseline, high-precision, broad-band photometry from missions like \emph{Kepler}, K2, and TESS.  These observations are crucial inputs for assessments of biosignatures in exoplanetary atmospheres and surface ultraviolet radiation dosages for habitable-zone planets around low-mass stars.   A limitation of these assessments, however, is the lack of near-ultraviolet spectral observations of stellar flares.  To motivate further empirical investigation, we use a grid of radiative-hydrodynamic simulations with an updated treatment of the pressure broadening of hydrogen lines to predict the $\lambda \approx 1800-3300$ \AA\ continuum flux during the rise and peak phases of a well-studied superflare from the dM3e star AD Leo. 
These predictions are based on semi-empirical superpositions of radiative flux spectra consisting of a high-flux electron beam simulation with a large, low-energy cutoff ($\gtrsim 85$ keV) and a lower-flux electron beam simulation with a smaller, low-energy cutoff ($\lesssim 40$ keV).  The two-component models comprehensively explain the hydrogen Balmer line broadening, the optical continuum color temperature, the Balmer jump strength, and the far-ultraviolet continuum strength and shape in the rise/peak phase of this flare.    We use spatially resolved analyses of solar flare data from the Interface Region Imaging Spectrograph, combined with the results of previous radiative-hydrodynamic modeling of the 2014 March 29 X1 solar ﬂare (SOL20140329T17:48), to interpret the two-component electron beam model as representing the spatial superposition of bright kernels and fainter ribbons over a larger area. 

\tiny
\keyFont{ \section{Keywords:} flares-stars, flares-sun, habitability and astrobiology, spectroscopy, near-ultraviolet continuum} 
\end{abstract}

\section{Introduction}
Rapidly rotating, magnetically active M dwarf (dMe) stars occasionally flare with energies that are factors of 100--10,000 greater than the most energetic solar flares that have been observed in the modern era.  These so-called ``superflares'' provide insight into the physics of extreme plasma conditions attained in stars \citep{Osten2007, Testa2008, Osten2010, Osten2016, Karmakar2017} and possibly also the young Sun \citep{Ayres2015, Maehara2015, Namekata2021}. Further afield, observations of these superflares are widely used in the characterization of the high-energy radiation environments in the habitable zones of low-mass stars \citep{Smith2004, Segura2010}, which are a primary target for exoplanet transit spectroscopy \citep[e.g.,][]{Scalo2007, Belu2011, deWit2018, Fauchez2019, JWST1, JWST2} with the \emph{James Webb Space Telescope} (JWST).  The JWST and future extremely large telescope facilities provide a means to determine whether exoplanets in or around the habitable zone of low-mass stars  retain atmospheres in the presence of high fluxes of stellar energetic particles and X-ray and extreme ultraviolet (XEUV) flare radiation.  The evolution of the atmosphere of Mars is thought to have undergone significant mass loss due to coronal mass ejections and XEUV heating from the Sun \citep{Jakosky2018}, which motivates investigations into the evolution of exoplanetary atmospheres that are much closer to stars that are highly magnetically active for billions of years \citep{West2008}.

\citet{Segura2010} simulated the impact of a superflare event on a non-magnetic, but otherwise Earth-like, exoplanet atmosphere  in the habitable zone of the dM3e flare star AD Leo.  They found that planetary ozone is largely depleted due to chemical reactions \citep[e.g.,][]{Scalo2007} that follow from incident scaled-up fluxes \citep{Belov2005} of solar energetic protons.  More recent simulations have considered the effects of repeated flaring and particle events after such a superflare has occurred \citep{Tilley2019, Howard2018}.  \citet{Howard2018} and \citet{Tilley2019} discuss the role of UV-C\footnote{According to the World Health Organization, the ultraviolet is comprised of three bands:  UV-C ($\lambda = 1000-2800$ \AA), UV-B ($\lambda = 2800 - 3150$ \AA), and UV-A ($\lambda = 3150-4000$ \AA).  We follow \citet{Abrevaya2020} and other recent studies of the biological impact of UV flares and denote the continuum radiation at $\lambda = 2000 - 2800$ \AA\ as UV-C; the ultraviolet radiation at $\lambda = 1000 - 2000$ \AA\ is also known as the very ultraviolet (VUV).} ($\lambda = 2000-2800$\AA), and in particular the wavelengths $\lambda = 2400-2800$ \AA, from repeated flaring in germicidal radiation surface fluxes  following an ozone-depletion event.  Recently, \citet{Abrevaya2020} conducted laboratory measurements of survival curves of microorganisms through exposure to a UV-C radiation flux inferred from optical observations of a superflare from the dM5.5e star Proxima Centauri \citep{Howard2018}. 

 The transient \citep{Loyd2018} and secular \citep{Venot2016} effects on ozone biosignature photochemistry caused by ultraviolet flares is an ongoing subject of research, especially in light of lack of direct observation of stellar energetic proton fluences and exoplanetary magnetic field properties \citep[see discussion in][]{Tilley2019}.  \citet{Loyd2018} demonstrate that assumptions of the ultraviolet continuum shape during flares can affect ozone photolysis rates \citep[see also][]{Howard2020}, while effects spanning several orders of magnitude on other important atmospheric constituents (CH$_4$, H$_2$O, O$_2$) are expected.   Significant effort has been invested into the empirical characterization of the quiescent and flaring spectra of low-mass stars in the far-ultraviolet wavelength region of $\lambda  = 1100-1800$ \AA\ through the MUSCLES, Mega-MUSCLES, and HAZMAT treasury programs with the \emph{Hubble Space Telescope} \citep[][and see also \citealt{Feinstein2022}]{Shkolnik2014, Loyd2014, Froning2019, Loyd2018Hazmat, France2020, Wilson2021}.   However, much still remains unknown about the spectral characteristics of transient, impulsive-phase, near-ultraviolet (NUV) enhancements in flare radiation from $\lambda \approx 2000-3300$ \AA\ \citep[e.g.,][]{Robinson2005, Hawley2007, Brasseur2019, Fleming2022}, which is thought to account for a large percentage ($\approx$25\%) of the $\lambda = 1200-8000$ \AA\ radiated energy \citep{HP91}. The largest X-ray solar flare of Sunspot Cycle 24 was recently studied in spatially integrated light at continuum wavelengths through a $\Delta \lambda \approx 300$ \AA\ bandpass around $\lambda \approx 2000$ \AA\ \citep{Dominique2018}.

Many studies of stellar flares have utilized photometry from high-precision missions of \emph{Kepler}, K2, and TESS, which observe through broad white-light bandpasses in the optical and near-infrared.  Photochemistry and habitability calculations often use extrapolations to shorter ultraviolet wavelengths by assuming a $T \approx 9000 - 10,000$ K blackbody spectrum \citep[e.g.,][]{Gunther2020}; this assumption has also been widely employed for calculations of bolometric energies in statistical analyses \citep[e.g.,][]{Shibayama2013, Yang2017}.  We refer the reader to \citet{Howard2020} and Brasseur et al.\ (2022, submitted to \emph{ApJ}) for discussions about several unprecedented problems raised by recent studies of multi-wavelength broadband photometry of stellar superflares.   

Very few stellar flare spectral observations exist with near-ultraviolet coverage and contemporaneous optical spectra  \citep{HP91, Robinson1993, Wargelin2017, Kowalski2019HST} that would facilitate detailed tests of and improvements upon blackbody modeling of flares.  However, ground-based spectra suggest that the NUV flare continuum has non-negligible contributions from Balmer continuum radiation \citep{Kowalski2010, Kowalski2013, Kowalski2016}.  Most recently, \citet{Kowalski2019HST} analyze  $\lambda \approx 2500-7400$ \AA\ flare spectra and photometry over two events from the dM4e star GJ 1243.  Detailed modeling demonstrates that a $T=9000$ K blackbody fit to the blue-optical continuum at $\lambda \approx 4000-4800$ \AA\ underpredicts the NUV flare flux by factors of $2-3$ during these two events; the discrepency was tied to a moderately-sized jump in the continuum flux around the Balmer limit, a confluence of Fe II emission lines through the NUV, and the bright Mg II $h$ and $k$ resonance lines \citep[see also][]{Hawley2007}.  These ``hybrid flare'' (HF) or ``gradual flare'' (GF) \citep[see][]{Kowalski2013} events exhibit the largest Balmer jumps that have been detected spectroscopically in dMe flares, and \citet{Kowalski2019HST} argue that other dMe events that are categorized as ``impulsive flare'' (IF) events according to their broadband time evolution, smaller Balmer jumps, and hotter blackbody fits to the optical  \citep{Kowalski2013} and $U$-band ($\lambda = 3260-3940$ \AA) continua \citep{Fuhrmeister2008} yet require spectroscopic investigation at shorter wavelengths.  The optical spectral properties of impulsive-type M dwarf flares are crucial in our understanding of fundamental flare physics because they are not reproduced in simulations with typical, solar-type electron beam \citep{Allred2006} or intense XEUV radiation fields \citep{HF92};  large continuum optical depths are required in the flare chromosphere \citep[e.g.,][]{Livshits1981, Kowalski2015} or photosphere.

In this paper, we present radiative-hydrodynamic model predictions of the NUV flare continuum during the Great Flare of AD Leo \citep{HP91}, a particularly well-studied, impulsive-type superflare.  The optical emission line data have previously been modeled in detail in \citet{HF92} using NLTE, X-ray backwarming calculations and in \citet{Allred2006} with electron beam heating with the \texttt{RADYN} code.  However, comprehensive models of the powerful optical continuum radiation and the broadening of the hydrogen Balmer line series have not yet been addressed.  The multi-wavelength spectra of this event have been widely utilized for empirically-driven models of exoplanet photochemistry and surface UV dosages \citep{Segura2010, Venot2016, Ranjan2017, Tilley2019, Estrela2020}.   

This paper is organized as follows.  In Section \ref{sec:observations}, the AD Leo Great Flare spectrum and photometry data that are used for model fitting are briefly reviewed.  Section \ref{sec:rhdmodels} describes the radiative-hydrodynamic flare models, which are fit to the H$\gamma$ emission line in the spectrum corresponding to the rise and peak phases of the Great Flare (Section \ref{sec:analysis}).  We calculate the Balmer line merging in the spectral region around the Balmer limit (Section \ref{sec:balmerlimit}) to further justify our two-component model fitting approach.   We consider the FUV spectrum during the early impulsive phase and independently fit to the observed continuum distribution of the Great Flare in order to make a new continuum model prediction for the NUV wavelength range that was not observed during this time (Section \ref{sec:broadbandfits}).  The interpretation of the results is discussed in Section \ref{sec:discussion} in terms of the spatial distribution of intensity in an image of a well-studied solar flare; we present new calculations for habitable zone UV-C fluxes in Section \ref{sec:hz}. We conclude in Section \ref{sec:conclusions}.

\section{Observations} \label{sec:observations}

\subsection{The 1985 April 12 Great Flare of AD Leo}

\begin{figure}
   \centering
       \includegraphics[width=0.5\textwidth]{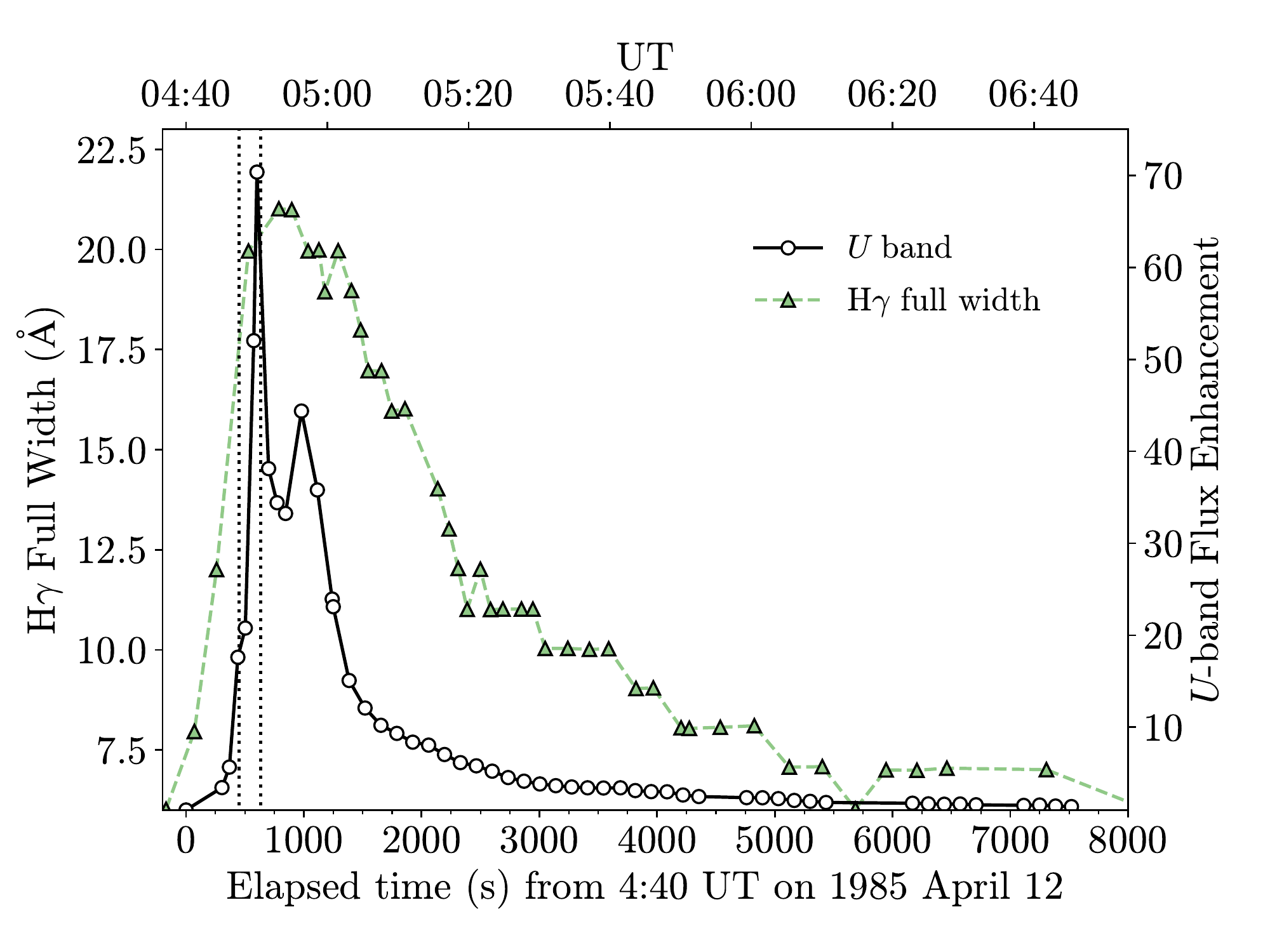}
       \caption{The $U$-band light curve (right axis) of the Great Flare of AD Leo and the full-width (left axis) of the H$\gamma$ emission line from \citet{HP91}.  The $U$-band flux is normalized to quiescence.  The start and stop times of the exposure corresponding to the rise/peak phase spectrum is indicated with two vertical dotted lines. }
       \label{fig:gflc}
   \end{figure}

The Great Flare of AD Leo was a large-amplitude, superflare event with an energy of nearly $10^{34}$ erg emitted in the $U$ band.   The available data are the broadband $UBVR$ photometry and the optical spectra covering $\lambda = 3560-4400$ \AA\ at a resolving power of $R \approx 1240$.   The exposure times of the spectra varied between one and three minutes \citep[see][for details]{HP91}.  Here, we model the spectrum that integrated over most of the rise and first peak, labeled as the ``542s'' spectrum in \citet{HP91} and ``S\# 36'' in the analysis of \citet{Kowalski2013}\footnote{\citet{Segura2010} refers to the spectrum at 915~s as the peak spectrum;  this is S\# 39 in the labeling scheme of \citet{Kowalski2013} and corresponds to the second, lower-amplitude peak in the impulsive phase.}.  The light curve of the $U$ band is shown in Figure \ref{fig:gflc} with the integration time of the rise/peak spectrum indicated. 

The Great Flare exhibits all of the spectral and light curve characteristics of a highly ``impulsive-type'' stellar flare, according to the ``IF'' classification in \citet{Kowalski2013}.   Specifically, the blue-optical spectra were fit with a color temperature of $T \approx 11,600$ K, and small Balmer jump in the $U$ band suggests that Balmer recombination radiation is important at shorter wavelengths \citep[Fig.\ 9 of ][]{Kowalski2013}.  During the early impulsive phase, a FUV spectrum (described below) constrains the peak of the continuum to the $U$ band with a turnover toward shorter wavelengths that was found to be most consistent with a $T = 8500-9500$ K blackbody among the models that were available at the time \citep{HF92}.
 \citet{HP91} analyzed the highly broadened, symmetric wings of the Balmer series, which were attributed to the Stark effect.  The full-width evolution of the hydrogen Balmer H$\gamma$ emission line from \citet{HP91} is reproduced in Figure \ref{fig:gflc}.  The rise/peak spectrum corresponds to the first observation that exhibits very broad H$\gamma$ wings.

The AD Leo Great Flare was observed with ultraviolet spectroscopy with the International Ultraviolet Explorer (IUE).  A FUV spectrum was observed in the short-wavelength channel (SWP; $\lambda = 1150-2000$ \AA), which integrated over the first 900~s of the flare (up until 4:55 UT) and included 41 minutes of quiescence.  The fluxes in the major emission lines and in the continuum longward of 1780 \AA\ saturated the detector.  NUV spectral observations in the long wavelength band (LWP; $\lambda = 1900-3100$ \AA) of IUE started at 5:00 UT, which is about midway during the second, fast decay phase in the $U$-band light curve in Figure \ref{fig:gflc}.  The LWP observation from 5:00--5:20 UT was split into five sub exposures, each $3-8$ minutes in duration (see Fig.\ 1 of \citet{HP91} for the $\lambda = 2000$ \AA\ and $2800$ \AA\ continuum flux evolution over the first set of five sub-exposures).  In the flare,  some emission lines in the NUV were saturated as well \citep[see ][for details about how these data have been utilized by interpolation and binning]{Segura2010}.   Since there were no NUV spectra covering the rise and peak phases, we do not consider the IUE/LWP spectra further in this study.  For detailed descriptions of the reduction and analyses of the IUE spectra, we refer the reader to \citet{HP91}.

\subsection{2014 Mar 29 X1 Solar Flare}

The 2014 March 29 GOES class X1 ﬂare (SOL20140329T17:48) is one of the best-observed and most-widely studied solar flares from Sunspot Cycle 24 \citep[e.g.,][]{Heinzel2014, Battaglia2015, Aschwanden2015, Young2015, Rubio2016, Woods2017, Rubio2017, Polito2018, Kleint2018, Zhu2019}.  The flare was observed by the Interface Region Imaging Spectrograph \citep[IRIS;][]{DePontieu2014} with FUV and and NUV longslit spectroscopy \citep[see, e.g.,][for detailed descriptions of these spectra]{Kleint2016, Kowalski2017Mar29}.  Narrow-band ($\Delta \lambda \sim 4$ \AA), slit jaw images (SJI) in the NUV at Mg II 2796 (SJI 2796) and at 2830 \AA\ (SJI 2832) are available for contextual information about the flare brightenings that cross the IRIS slit. 
For this study, a level-2, SJI 2832 image at 17:46 UTC is retrieved from the IRIS data archive hosted at the Lockheed Martin Solar and Astrophysical Laboratory\footnote{\url{https://iris.lmsal.com/search/}}.
The SJI 2832 image corresponds to the early impulsive phase of the hard X-rays at $E\ge 25$ keV and has been analyzed in \citet{Kowalski2017Mar29}.  The IRIS slit location stepped through the ribbons in this flare, resulting in a cadence of 75~s for the raster and SJI 2832 images.   Following previous analyses \citep[e.g.,][]{Kowalski2017Mar29}, we convert the level-2 data in units of DN s$^{-1}$ pix$^{-1}$ to an equivalent, constant intensity value, $\langle I_{\lambda} \rangle_{\rm{SJI}}$, over the SJI bandpass using the time-dependent effective area curves \citep{Wulser2018} provided by the IRIS mission. 

X-ray imaging data from the Reuven Ramaty High-Energy Solar Spectroscopic Imager \citep[RHESSI;][]{Lin2002} were retrieved from the new RHESSI data archive\footnote{\url{https://hesperia.gsfc.nasa.gov/rhessi_extras/flare_images/image_archive_guide.html}}.  We use the 6--12 keV and 50--100 keV imaging over the time-interval of 17:45:58.7 to 17:46:32.0.  We refer the reader to  \citet{Battaglia2015} and \citet{Kleint2016} for higher temporal and spatial resolution analyses of the RHESSI data for the 2014 March 29 flare.

\section{Radiative-Hydrodynamic Flare Models} \label{sec:rhdmodels}
To model the rise/peak phase spectrum of the AD Leo flare, we use results from a grid of radiative-hydrodynamic (RHD) models calculated with the \texttt{RADYN} code \citep{Carlsson1992B, Carlsson1995, Carlsson1997, Carlsson2002, Allred2015}.  All of the details about the simulation setup will be described in a separate paper (Kowalski et al.\ in preparation),  but a brief summary is presented here.  To simulate flare heating, we model the energy deposition from a power-law distribution (hereafter, ``beam'') of electrons, which is calculated in a 1D magnetic loop of half-length $10^9$ cm, a constant surface gravity of log$_{10}$\ $g/[$cm s$^{-2}]=4.75$, and a uniform cross-sectional area.  The effective temperature of the starting atmosphere is $T_{\rm{eff}} \approx 3600$ K \citep[see the Appendix of][for details regarding the starting atmosphere]{Kowalski2017Broadening}.  The equations of mass, momentum, internal energy,  and charge are solved on an adaptive grid with the equations of radiative transfer and level populations for hydrogen, helium, and Ca II.   The electron beam is injected at the loop apex with a ramping flux to a maximum value at $t=1$~s, followed by a decrease until $t=10$~s according to the pulsed injection profile prescription in \citet{Aschwanden2004}. The heating rate as a function of depth is determined by the steady-state solution to the Fokker-Planck equation for energy loss and pitch angle scattering due to Coulomb collisions using the module that was further developed into the \texttt{FP} code \citep{Allred2020}. The heating rate is recalculated at every time-step in the radiative-hydrodynamic simulation \citep[see][]{Allred2015}.  In this first generation of models, return current electric fields and magnetic mirroring forces \citep{Allred2020} are not considered.  However, hydrogen Balmer line spectra (H$\alpha$, H$\beta$, H$\gamma$) are properly modeled using the Doppler-convolved, TB09+HM88 line profile functions \citep{Smith1969, Vidal1970, Vidal1971, Vidal1973, HM88, Tremblay2009}, which accurately capture the pressure broadening from ambient, thermal electrons and ions in the density regimes of flare chromospheres \citep{Kowalski2017Broadening, Kowalski2022}. 

Stellar flare hard X-ray emission is below current detection limits, except in the most energetic events \citep[e.g.,][]{Osten2007, Osten2016}, and millimeter / radio observations at optically thin frequencies have been reported only recently \citep{MacGregor2020}.  The paucity of direct constraints on accelerated electrons in stellar flares thus necessitates a grid of models covering a large parameter space of electron beam heating.
Our grid of M dwarf flare models includes a large range of low-energy cutoff ($E_{\rm{c}}$) values:  $17, 25, 37, 85, 150, 200, 350, $ and $500$ keV.   All of the selected models from the grid are calculated for injected electron beam number fluxes with hard power-law indices of $\delta = 2.5 - 4$, which are consistent with available stellar flare constraints \citep{Osten2007, MacGregor2018, MacGregor2020, MacGregor2021}.  The peak injected beam energy flux densities (hereafter, ``flux'') span  four orders of magnitude: $10^{10}$ (F10), $10^{11}$ (F11), $10^{12}$ (F12), and $10^{13}$ (F13) erg cm$^{-2}$ s$^{-1}$.  For this study, we select five models with maximum (``m'') injected beam fluxes of $10^{13}$ erg cm$^{-2}$ s$^{-1}$ (``mF13''), low-energy cutoffs of $E_c=37, 85, 150, 200,$ and $ 500$ keV, and a power-law index of $\delta = 3$;  these are referred to as the mF13-37-3, mF13-85-3, mF13-150-3, mF13-200-3, and mF13-500-3 models, respectively\footnote{A corresponding grid of models is calculated using a constant beam flux injection; these models are indicated with a ``c''-prefix, such as cF13-85-3.  } These models are especially notable because they reproduce $T \approx 10,000$ K color temperatures in the blue-optical wavelength range  and small Balmer jump ratios, as reported in many M dwarf flare spectral observations \citep[e.g.,][]{Mochnacki1980, Fuhrmeister2008, Kowalski2013, Kowalski2016}.  The justification for selecting these high-flux models will be discussed further in Section (\ref{sec:fits}). 

The mF13-37-3 model is a recalculation of the \texttt{RADYN} simulation from \citet{Kowalski2016} (see also \citet{Kowalski2015}) with a pulsed beam flux injection.  The atmosphere in the new model follows a similar evolution with the development  a dense ($n_e = 5 \times 10^{15}$ cm$^{-3}$), cool chromospheric condensation at $t \approx 2.2$~s.   The mF13-85-3 and mF13-150-3 models produce relatively small amounts of coronal heating and relatively fast upflows ($\approx 5-20$ km s$^{-1}$) in the flare chromosphere because most of the beam energy is deposited into the deep chromosphere.  Without magnetic mirroring and return current electric field forces, long-lasting chromospheric condensations do not develop as in the mF13-37-3 model.  However, large ambient electron densities ($n_e \approx 1-7\times10^{15}$ cm$^{-3}$) are attained due to thermal ionization of hydrogen by the beam heating in low-lying chromospheric layers \citep[see the Appendices of][for a description of several, similar large, low-energy-cutoff models]{Kowalski2017Broadening}.  These charge densities refer to the atmospheric (ambient/thermal) proton and electron densities that pressure broaden the hydrogen lines that we model in Sections \ref{sec:fits} --  \ref{sec:balmerlimit}.  The nonthermal electron densities are many orders of magnitude smaller in the chromosphere.  The large, low-energy-cutoff models represent a semi-empirical approach in the spirit of the static flare atmospheres of \citet{Cram1982}, but the \texttt{RADYN} models include time-dependent atmospheric thermodynamics that are calculated self-consistently with beam heating.

In addition to the F13 models, a lower beam flux that has been used to model IRIS NUV spectra of a solar flare  \citep{Kowalski2017Mar29} and the broadening of the hydrogen Balmer series \citep{Kowalski2022} has been injected into our M dwarf atmosphere for a duration of $15$~s.  The electron beam parameters ($E_c = 25$ keV, $\delta = 4$, flux of $5\times10^{11}$ erg cm$^{-2}$ s$^{-1}$) for this model (``c15s-5F11-25-4'') were selected to be consistent with those that were inferred through the collisional thick target modeling of RHESSI hard X-ray data of the 2014 Mar 29 solar flare \citep{Kleint2016}.   Similar to the analogous simulation in the solar atmosphere, a dense chromospheric condensation develops by $t \approx 4$~s with densities of $n_e \approx 5\times 10^{14}$ cm$^{-3}$.  We also calculate a model (m5F11-25-4) with a shorter, pulsed injection profile in the same form as for the pulsed F13 beams.  Several other models that are considered in this work are two intermediate flux models (mF12-37-3 and m2F12-37-2.5) with hard power-law distributions ($\delta = 2.5$ and $3$) and intermediate low-energy cutoff values ($E_c = 37$ keV).  A similar model to the mF12-37-3 beam was analyzed in \citet{Namekata2020}, who found satisfactory agreement between the broadening of the hydrogen Balmer $\alpha$ emission line in the model and in the observation of a superflare event from AD Leo. 

The parameters of the RHD models that are used in the remainder of this work are summarized in Table \ref{table:models}.

\clearpage

\begin{landscape}

\begin{table}[h!]
\begin{tabular}{ ||p{3cm}|p{1.75cm}|p{1cm}|p{1cm}|p{1cm}|p{1cm}|p{2cm}|p{2cm}|p{1.25cm}|p{1.25cm}|  }
 \hline
 \multicolumn{10}{|c|}{Table 1 -- \texttt{RADYN} Electron Beam Heating Models} \\
 \hline
 \hline
 \renewcommand{\arraystretch}{1.5}
 Model &
Beam Flux &
$t_{1/2}$ &
$t_{\rm{end}}$ &
$E_c$ &
$\delta$ &
C4170$^{\prime}$ & $F^{\prime}_{\rm{H}\gamma}$ & $F^{\prime}_{\rm{H}\gamma} \div $ \newline C4170$^{\prime}$ & H$\gamma$ Eff.\ \newline Width \\
\hline
 &  & (s) & (s) & (keV) &  & (erg cm$^{-2}$ s$^{-1}$ \AA$^{-1}$) & (erg cm$^{-2}$ s$^{-1}$) & (\AA) & (\AA) \\
 \hline
 \hline
mF13-85-3 & max F13 & 2.3 & 10 & 85 & 3 &  $1.14 \times10^8$ & $1.62 \times10^9$ & $14.2 $ & $9.1$ \\
mF13-150-3 & max F13 & 2.3 & 10 & 150 & 3 &  $1.54 \times10^8$  &  $1.16\times10^9$  & $7.5 $ & $10.3$ \\
mF13-200-3 & max F13 & 2.3 & 10 & 200 & 3 & $1.73 \times10^8$  & $6.21 \times10^8$ & $3.6$ & $10.4$  \\
mF13-500-3 & max F13 & 2.3 & 10 & 500 & 3 &  $1.87 \times10^7$  & $-1.37 \times10^9$ & $-107$ & $-47.7$  \\
mF12-37-3 & max F12 & 2.3 & 10 & 37 & 3 & $3.51 \times 10^6$ & $5.45 \times10^8$  & $155.2$ & $3.8$ \\
m2F12-37-2.5 & max 2F12 & 2.3 & 10 & 37 & 2.5 & $1.72 \times10^7$  & $8.65 \times10^8$  & $50.3$ & $5.8$ \\
mF13-37-3 & max F13 & 2.3 & 10 & 37 & 3 &  $7.94 \times10^7$  & $2.22 \times10^9$ & $27.9$ & $12.1$ \\
m5F11-25-4 & max 5F11 & 2.3 & 10 & 25 & 4 & $6.35 \times10^5$ & $2.95 \times10^8 $  & $464.6$ & $2.2$ \\
c15s-5F11-25-4 & const 5F11 & 15 & 15 & 25 & 4 & $3.30 \times10^6$  & $1.33 \times10^9$  & $402.4$ & $5.3$  \\
 \hline 
\end{tabular}  
\caption{  $t_{1/2}$ is the full-width-at-half-maximum of the injected beam heating pulse; $t_{\rm{end}}$ indicates the end of the simulation and the duration over which the temporal averages of the model spectra are calculated.  The effective width of H$\gamma$ is defined as the integral of the continuum-subtracted, peak-normalized emission line profile \citep{Kowalski2022}; note that the H$\gamma$ profile is an absorption profile in the mF13-500-3 model. }
\label{table:models}
\end{table}

\end{landscape}

\clearpage

\section{Model Spectrum Analysis} \label{sec:analysis}
We leverage the new hydrogen pressure broadening profiles that have been incorporated into \texttt{RADYN} to examine the models that reproduce the Balmer jump strength and blue-optical continuum color temperature.   The Balmer H$\gamma$ emission line broadening and nearby blue-optical continuum fluxes of the Great Flare are the focus of our modeling analyses (Sections \ref{sec:ratios}-\ref{sec:fits}).  In Section \ref{sec:balmerlimit}, we extend the detailed calculations to spectra of the entire Balmer line series.

  \subsection{Average Model Line-to-Continuum Ratios} \label{sec:ratios}
 We first describe a simple method that allows comparisons of 1D loop models to the Great Flare spectra, which  are not spatially resolved.  Over an exposure time of 180~s, we reasonably expect that many sequentially ignited, spatially distinct, $\Delta t = 10$~s pulses accumulate flare radiation in the spatially unresolved, observed flare spectrum.  For each RHD model, we thus calculate a coadded spectrum from the radiative surface flux spectra at every $\Delta t = 0.2$~s by temporal averaging over the duration given by $t_{\rm{end}}$ in Table \ref{table:models}.  These coadded spectra are used in all analyses, unless otherwise indicated.

\begin{figure}
\begin{centering}
       \includegraphics[width=0.65\textwidth]{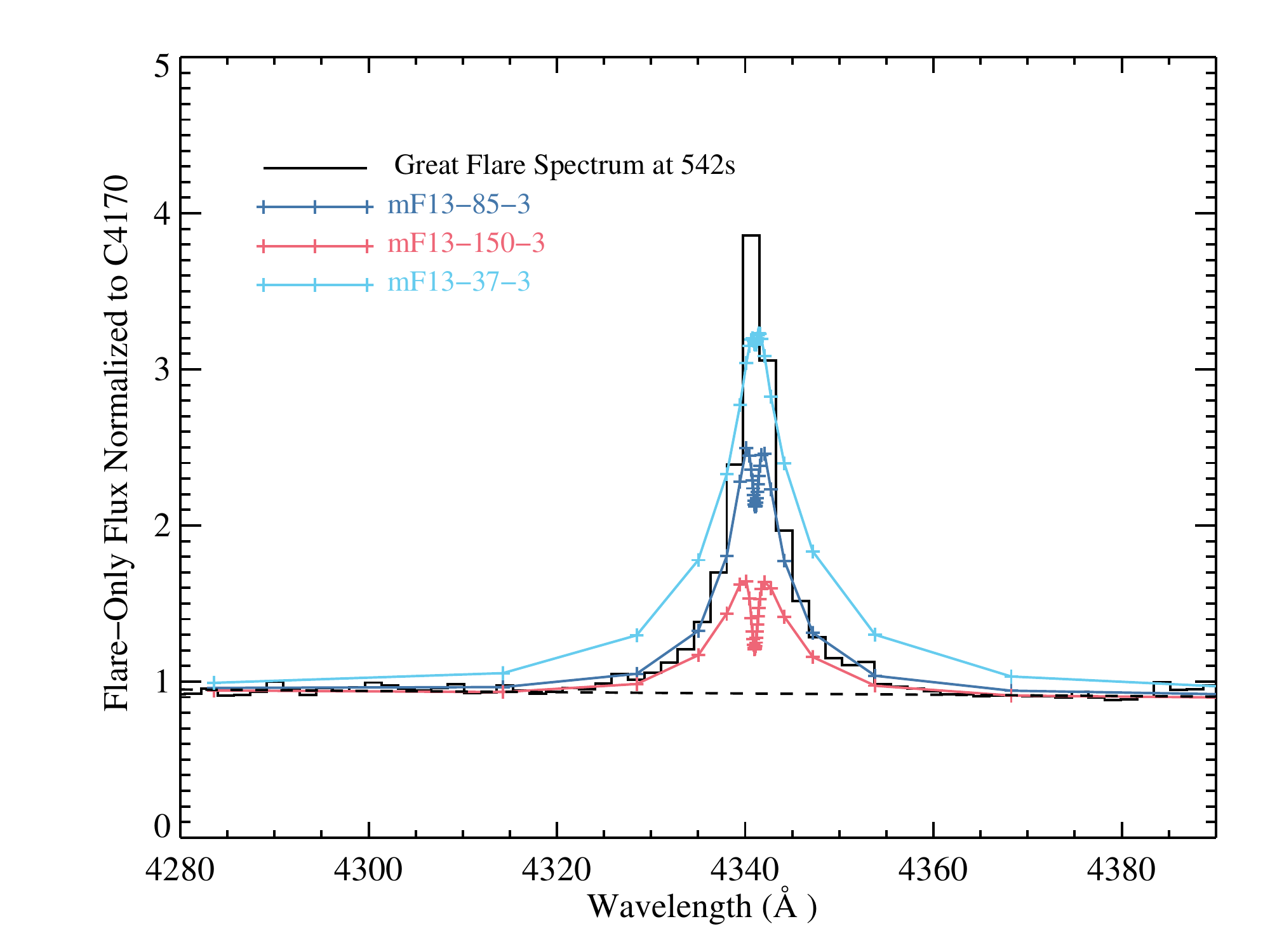}
       \caption{ Comparisons of several F13 model spectra of H$\gamma$ directly from \texttt{RADYN} (and thus have relatively coarse wavelength sampling in the far wings) to the Great Flare rise/peak phase spectrum.  Each model has been scaled to the observed continuum flux, C4170$^{\prime}$.  The mF13-37-3 model prediction is far too broad, while the mF13-85-3 and mF13-150-3 models do not exhibit an amount of broadening that exceeds the observation in the line wings.  The dashed line shows a detailed continuum spectrum that is interpolated to the wavelengths over the H$\gamma$ line. }
       \label{fig:coarse_F13}
       \end{centering}
   \end{figure}

   Several coadded F13 model spectra around the H$\gamma$ emission line are normalized to the observed continuum flare-only flux  averaged over $\lambda = 4155-4185$ \AA\ in the Great Flare (Figure \ref{fig:coarse_F13}).  With an older, less accurate prescription of hydrogen line pressure broadening, \citet{Kowalski2015} found that a coadded F13 model with a double power-law beam distribution and $E_c = 37$ keV was an adequate model of the early/mid rise phase of a giant flare from the dM4.5e star YZ CMi.  As Figure \ref{fig:coarse_F13} clearly demonstrates, the mF13-37-3 model profile with the updated hydrogen broadening is far too broad even though times when the chromospheric condensation is not extremely dense and the emission lines are relatively narrow are included in the coadd.  The coadded mF13 spectra from the models with large, low-energy-cutoffs ($E_c = 85-150$ keV) adequately account for some or all of the flux in the H$\gamma$ wings, but these models of deep flare heating vastly under-predict the relative H$\gamma$ line-peak flux.

To quantitatively assess the models, we calculate several quantities from the detailed H$\gamma$ line profiles and the continuum spectra. Specifically, we calculate the continuum-subtracted, preflare-subtracted, line-integrated flux over the H$\gamma$ emission line (hereafter, $F^{\prime}_{\rm{H}\gamma}$), the preflare-subtracted  flux\footnote{Following traditional use, we denote flare-only quantities with a prime-symbol, and we refer to an observed spectral/monochromatic/specific flux density at Earth as the ``flux'';  we use ``spectral luminosity'' to refer to the luminosity per unit wavelength.} at $\lambda = 4170$ \AA\ (hereafter, C4170$^{\prime}$), and the ratio of these quantities (Table \ref{table:models}).  The 5F11 and F12 models produce ratios that are far too large compared to  $F^{\prime}_{\rm{H}\gamma}$/C4170$^{\prime} \approx 20$  that is calculated from the Great Flare observation \citep{Kowalski2013}, while the ratios from the F13 models with large low-energy cutoffs are too small.  This motivates linear superpositions of two RHD model components -- one high-flux (F13) component and one lower-flux (5F11 or F12) component -- to comprehensively reproduce the observed line-to-continuum ratios, the shape of the H$\gamma$ wing broadening, and the H$\gamma$ emission line peak flux. 

  \subsection{Two-component Model Fits to H$\gamma$ Line in the AD Leo Great Flare} \label{sec:fits}

A linear superposition of a high flux (F13), large low-energy cutoff beam component and a lower flux (5F11, F12, or 2F12), smaller low-energy cutoff  beam component is represented by the equation (Eq.\ \ref{eq:superpos}),

\begin{equation} \label{eq:superpos}
f^{\prime}_{\lambda, \rm{Model}} =  \left( X_{\rm{F13}}\,  S^{\prime}_{\lambda, \rm{F13}} + X_{\rm{5F11}}\, S^{\prime}_{\lambda, \rm{5F11}} \right) \frac{R^2_{\rm{star}}}{d^2}
\end{equation}

\noindent where $f^{\prime}_{\lambda, \rm{Model}}$ is the model flux at Earth.  The filling factor, $X$, is the exposure-time-averaged fraction of the visible hemisphere of the star that is flaring with the temporal coadd of the radiative surface flux spectrum, $S_{\lambda}$, calculated from each RHD model component.   $R_{\rm{star}} = 3 \times 10^{10}$ cm is the radius of AD Leo, and $d = 1.5 \times10^{19}$ cm is the distance to the star.  The preflare model surface flux spectrum is subtracted to give  the flare-only, model surface flux spectra,  $S^{\prime}_{\lambda} = S_{\lambda} - S_{\lambda, o}$, in Equation \ref{eq:superpos}.  To mitigate systematic errors in the far wings of the H$\gamma$ line (Fig.\ \ref{fig:coarse_F13}), which are coarsely sampled at 31 wavelength points in the \texttt{RADYN} calculation, we recalculate\footnote{All analyses have been performed on both the original 31-wavelength array and the 327-wavelength array.} the emergent surface flux spectra using a Feautrier solver on a 327 point wavelength grid with the frequency-independent, non-LTE source function from \texttt{RADYN} and a four-point, third-order interpolation of the line profile opacity from the Appendix of \citet{Vidal1973}.  The emergent radiative flux spectra of H$\gamma$ are time-averaged over the duration of each simluation, convolved with a Gaussian with a full-width-at-half maximum that corresponds to the instrumental resolution of 3.5 \AA, and binned to the wavelengths of the Great Flare spectra.

\begin{figure}
\begin{centering}
       \includegraphics[width=0.45\textwidth]{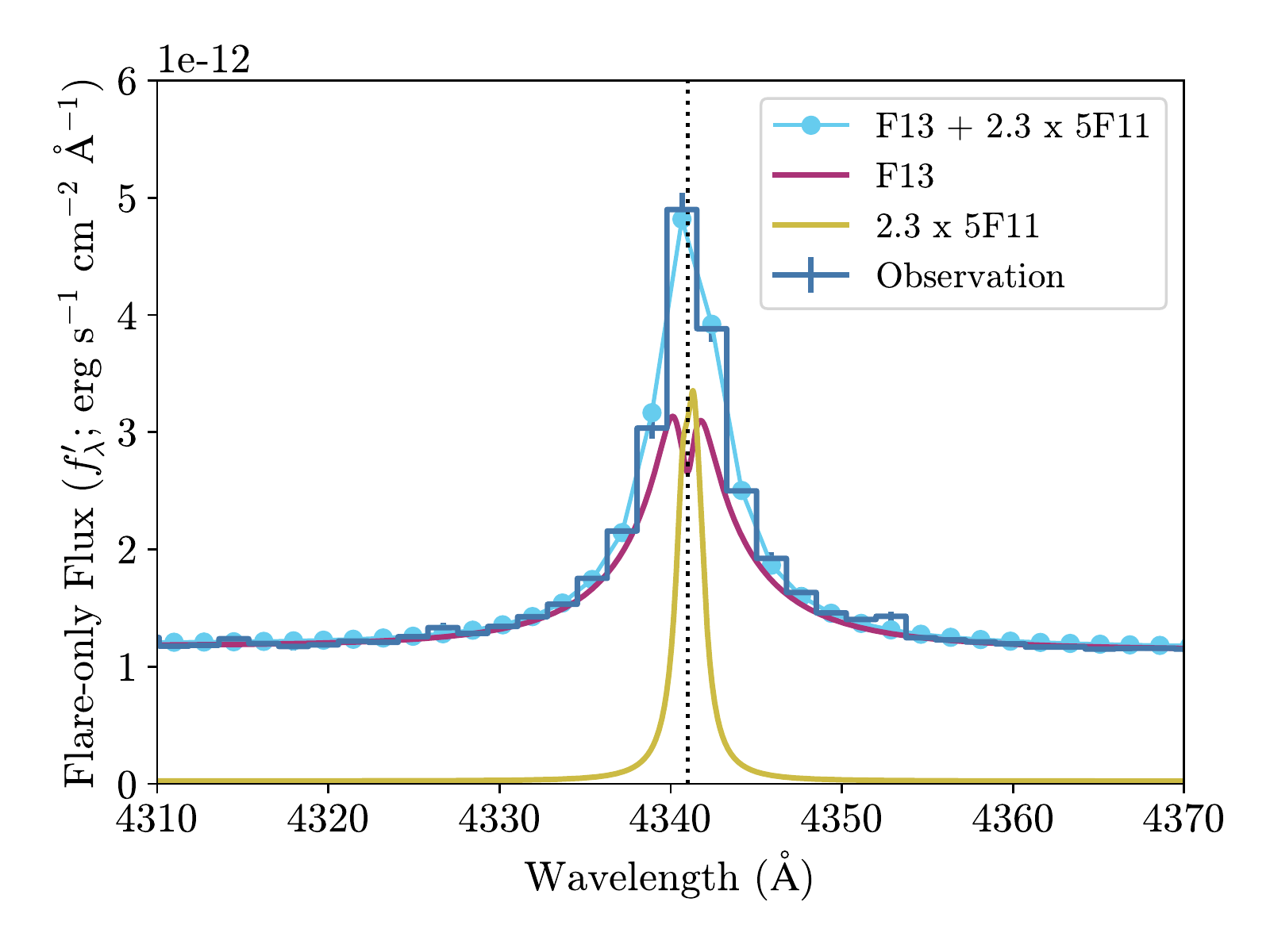}
       \includegraphics[width=0.45\textwidth]{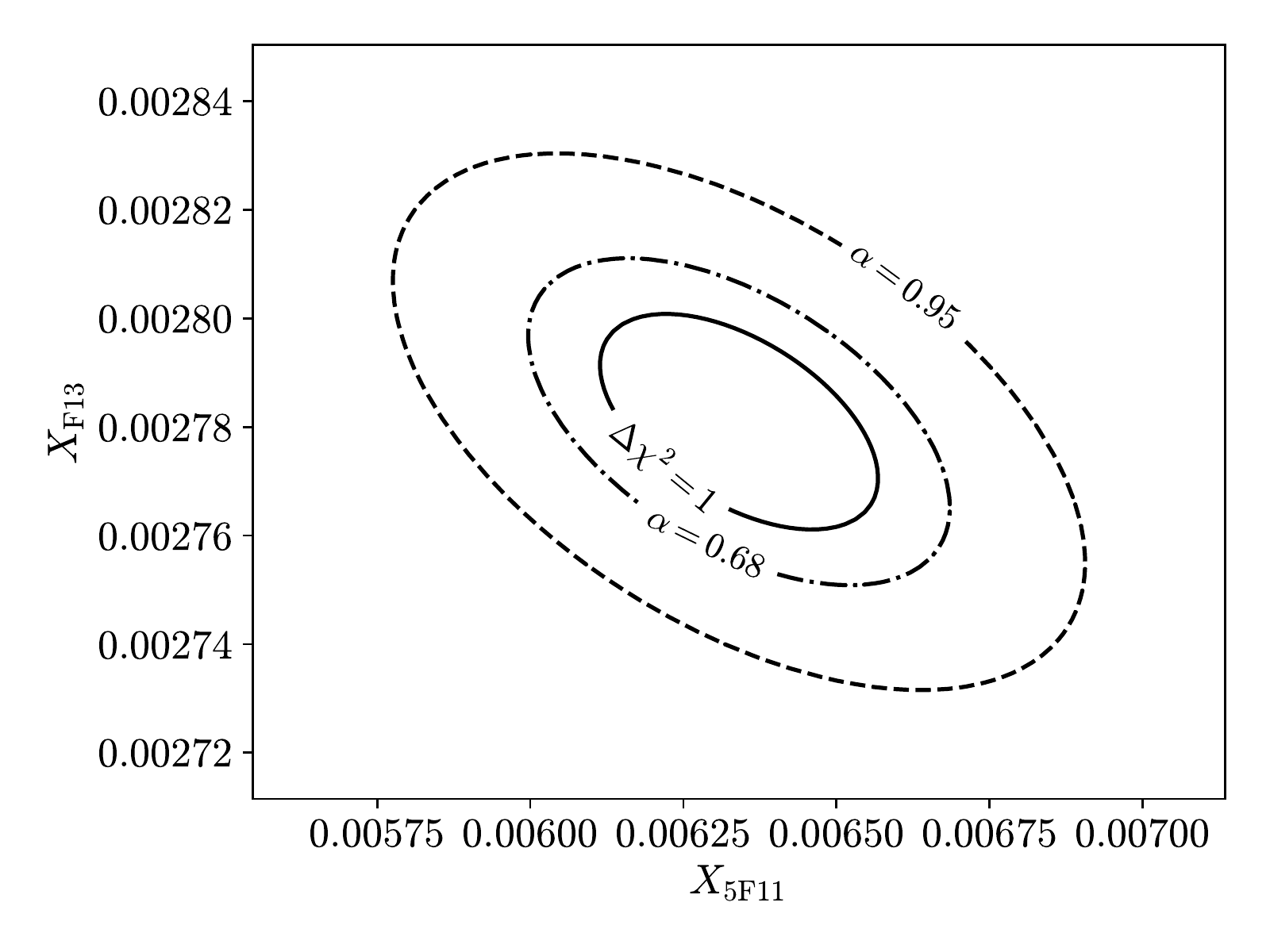}
       \caption{ (\textbf{Left}) A representative example of a satisfactory, two-component (mF13-85-3, m5F11-25-4) RHD model fit to the observed H$\gamma$ line profile and nearby continuum flux in the Great Flare of AD Leo.   (\textbf{Right}) Constant joint-likelihood contours for the model fit in the left panel. The maximum-likelihood estimates of the parameters and $1\sigma$, Gaussian marginal uncertainties are $\hat{X}_{\rm{F13,ML}} = 2.78 \times 10^{-3} \pm 2\times 10^{-5}$ and $\hat{X}_{\rm{5F11,ML}} = 6.34 \times 10^{-3} \pm 2.3\times 10^{-4}$  with a correlation coefficient of $\rho = -0.53$.  }
       \label{fig:example_fit}
       \end{centering}
   \end{figure}

We perform an inverse-variance-weighted, linear least-squares fit of the two parameters $X_{\rm{F13}}$ and $X_{\rm{5F11}}$ to the observed spectrum around the H$\gamma$ line.   The model surface flux spectra are the basis functions in the $n\lambda \times 2$ design matrix, $\Lambda$. The maximum likelihood (ML) estimates of the parameters are given by the standard matrix equation (Eq.\ \ref{eq:ML}),

\begin{equation} \label{eq:ML}
\hat{\vec{X}}_{\rm{ML}} = \begin{pmatrix} \hat{X}_{\rm{F13}, \rm{ML}} \\ \hat{X}_{\rm{5F11}, \rm{ML}}  \end{pmatrix}  = (\Lambda^{\rm{T}} C_{\vec{f}^{\prime}_{\lambda}}^{-1} \Lambda)^{-1}(\Lambda^{\rm{T}}C_{\vec{f}^{\prime}_{\lambda}}^{-1}\vec{f}^{\prime}_{\lambda}) 
\end{equation}

\noindent where $\vec{f}^{\prime}_{\lambda}$ is the observed $n\lambda \times 1$ flare-only flux at Earth (hereafter dropping the vector notation) as a function of wavelength and the $n\lambda \times n\lambda$ covariance matrix $C$ is populated with independent, Gaussian uncertainties, which are estimated from the data, $f^{\prime}_{\lambda}$.  The wavelength range from $\lambda = 4320 - 4361$ \AA\ is used in the fits, which are performed for all combinations of two models from the grid.  The models in Table \ref{table:models} are among the combinations with the lowest values of $\chi^2$ and were thus chosen as the focus of this study. 
Figure \ref{fig:example_fit} shows the result of one of the best fits with $\chi_{\rm{dof}}^2 = 1.2$ for $21$ degrees of freedom (dof). The quality of this fit is representative of many such results with two RHD component spectra consisting of a high-flux, large low-energy cutoff beam and a lower-flux, smaller low-energy cutoff beam.
In the right panel of Figure \ref{fig:example_fit}, we show likelihood contours for this fit to visualize typical uncertainties on the maximum-likelihood  estimates of the parameters.  The comparisons of the inferred filling factors of the high-flux (e.g., F13) and lower-flux (e.g., 5F11) models in each fit will be more useful in comparison to solar flare data (Section \ref{sec:interpretation}), and therefore we report values of $X_{\rm{rel}} = \frac{X_{\rm{5F11}}}{X_{\rm{F13}}}$.  For the model combination in Figure \ref{fig:example_fit}, the best-fit parameters and standard error propagation give $X_{\rm{rel}} = 2.28 \pm 0.08$.  The value of $F^{\prime}_{\rm{H}\gamma}/$C4170$^{\prime} = 19.9$ is remarkably consistent with this measured quantity from the observed flare spectrum.

The results for several representative combinations of models in Table \ref{table:models} with small values of $\chi^2$ are presented in Table \ref{table:chisqr}, indicating a range of value of $X_{\rm{rel}} \approx 1.5-10$.  Since these fits include only the H$\gamma$ line data for this flare, the relatively small differences in the various $\chi^2$ values in Table \ref{table:chisqr} are not strictly indicative of a global minimum.  The vast majority of all model combinations from the entire grid, however, result in $\chi^2$ values far in excess of those shown in Table \ref{table:chisqr}.  We think that an exploratory approach to the model grid predictions is more productive than an effort to find one model that best satisfies all constraints from the data, given the many assumptions in the RHD modeling (e.g., specific choice of low-energy cutoff values in the grid, assumptions of constant-area loop geometry and a constant power-law index value over each pulse).  The small values of $\chi^2$ are thus most informative for limiting the vast parameter space for further comparisons of our general modeling paradigm to the multi-wavelength data of the Great Flare.

\begin{table}[h!]
\begin{tabular}{ ||p{3cm}|p{3cm}||p{3cm}|p{3cm}||p{3cm}||  }
 \hline
 \multicolumn{5}{|c|}{Table 2 -- Least-Squares Fitting Results for H$\gamma$} \\
 \hline
 \hline
F13 Model & $X_{\rm{F13}}$ &  Lower-Flux Model  & $X_{\rm{rel}}$ &
$\chi^2_{\rm{dof}}$  \\
 \hline
mF13-85-3 & 0.0028 & m5F11-25-4 & 2.3 & 1.2 \\
mF13-150-3 & 0.0022 & m5F11-25-4 & 4.5 & 4.1   \\
mF13-150-3 & 0.0019 & c15s5F11-25-4 & 1.6 & 7.9  \\
mF13-200-3 & 0.0017 & mF12-37-3 & 4.9 & 1.9 \\
mF13-500-3 & 0.0009 & m2F12-37-2.5 & 9.6 & 1.7 \\
 \hline 
\end{tabular}  
\caption{  Least-squares fitting results for the H$\gamma$ emission line in the Great Flare of AD Leo.  }
\label{table:chisqr}
\end{table}

  \subsection{Models of the Balmer Limit}  \label{sec:balmerlimit}
In order to robustly extrapolate the models of this flare to the NUV wavelength range that was not observed during most of the impulsive phase, the Balmer jump strength in the observation should be satisfactorily reproduced.   However, many linear combinations of two models in the \texttt{RADYN} flare grid produce small Balmer jump ratios that are consistent with the measured range \citep[$\chi_{\rm{flare}} \approx 1.7-1.9$;][]{Kowalski2013} from the spectrum of the Great Flare.  For supplementary constraints, we compare the details of the merging of the Balmer series at $\lambda = 3646 - 4000$ \AA.  The last visible Balmer emission line is often used as an indication of the electron density, and \citet{HP91} discusses that the Balmer lines up to and including H15 or H16 are resolved in the Great Flare spectra.  Thus, our RHD model combinations should reproduce this salient property.

\begin{figure}
\begin{centering}
       \includegraphics[width=0.6\textwidth]{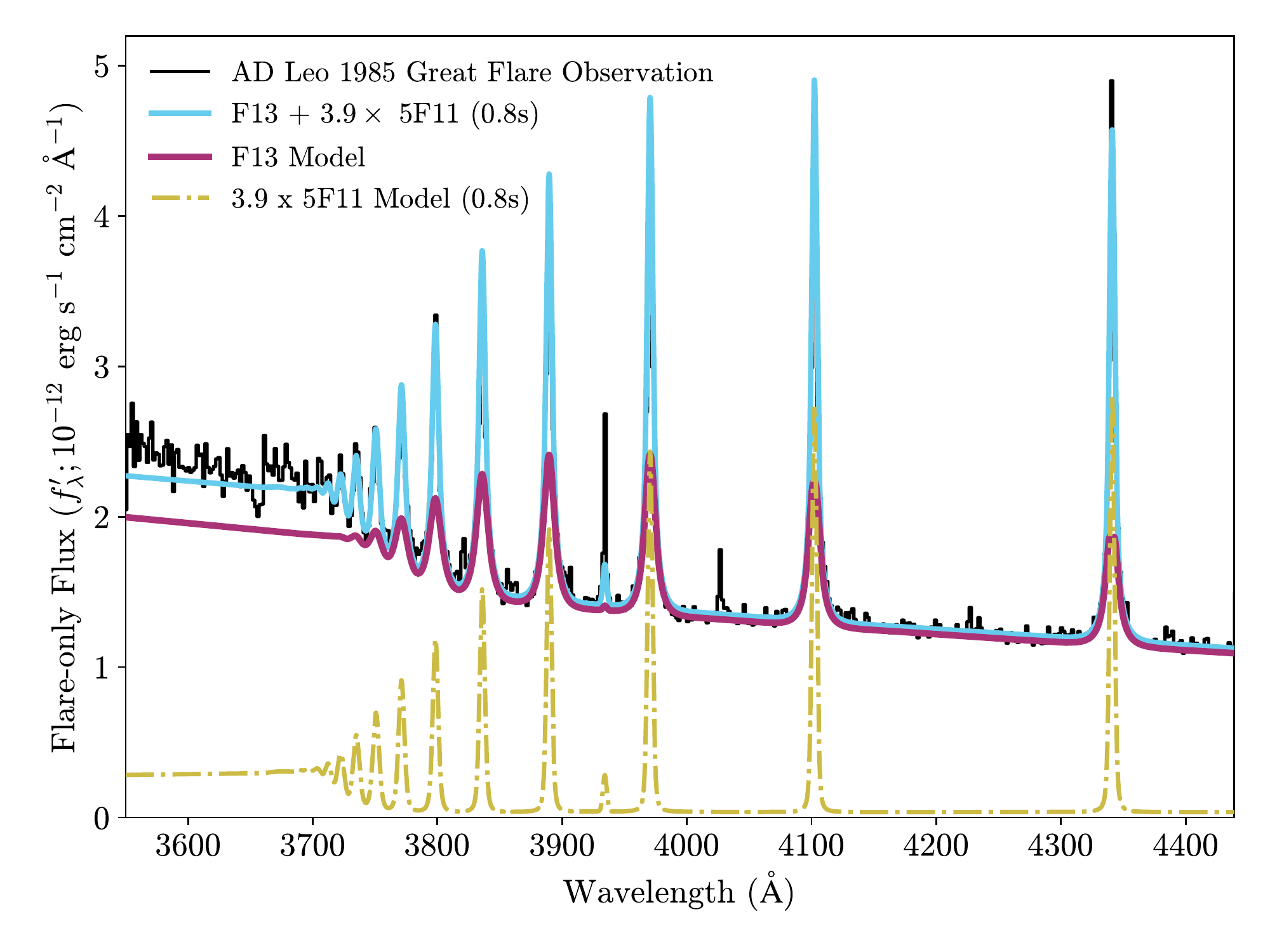}
         \includegraphics[width=0.6\textwidth]{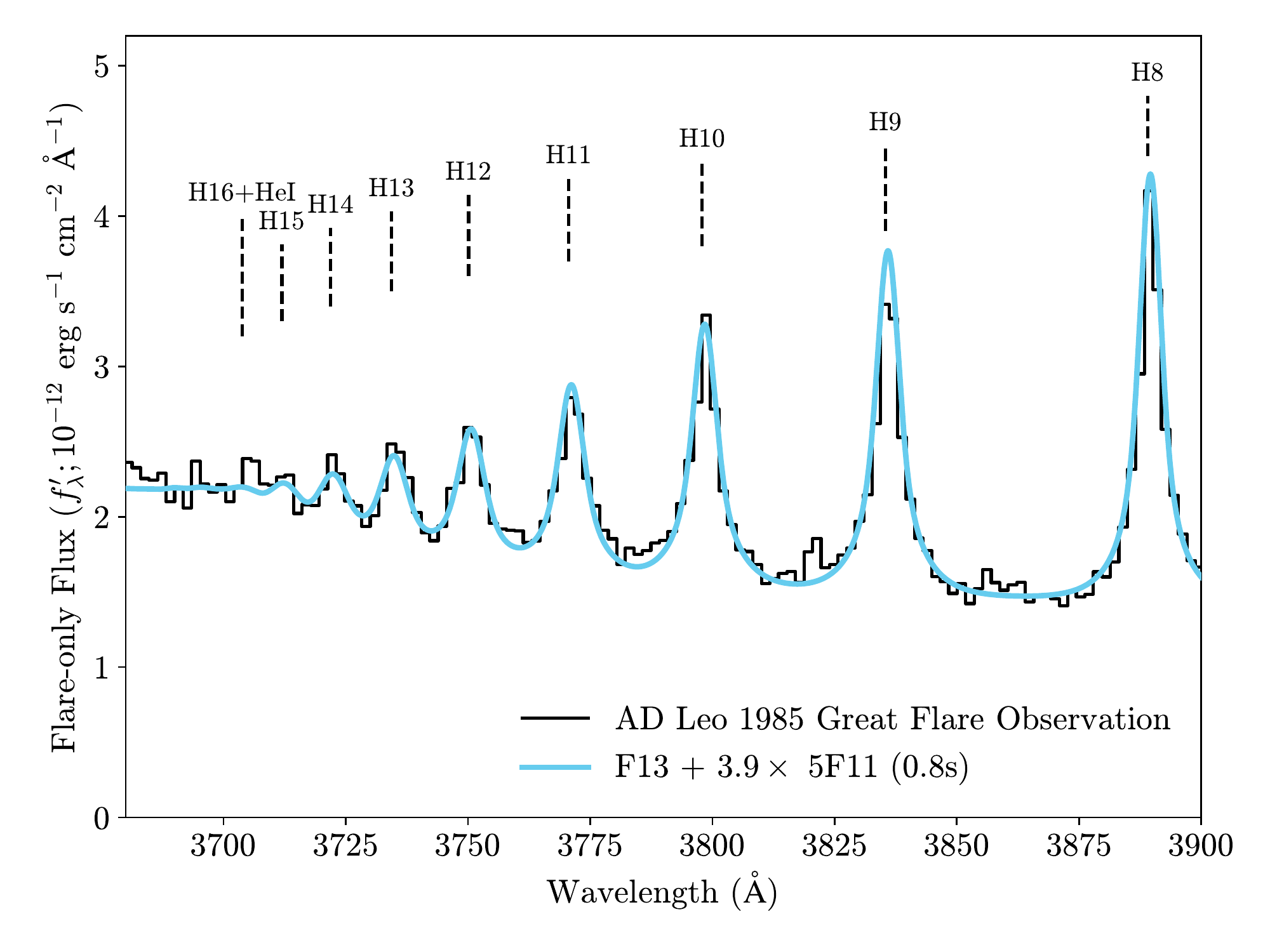}

       \caption{(\textbf{Top}) RH calculations with non-ideal opacity effects at the Balmer limit.  A linear superposition of the mF13-150-3 model, averaged over its duration, and the m5F11-25-4 model at $t=0.8$~s is shown for the combination that is constrained by a value of $F^{\prime}_{\rm{H}\gamma} / \rm{C4170}^{\prime}$ that is measured from the observed spectrum.   The individual component model calculations are shown: the m5F11 contributes to the narrow hydrogen line flux, the highest order Balmer lines, and  optically thin Balmer continuum flux in addition with the optically thick Balmer continuum flux from the F13.  The mF13 model accounts for nearly all of the optical continuum flux and far wing radiation. (\textbf{Bottom}) The pseudo-continuum from the merging of the Balmer H8 -- H15 line wings, the dissolved level continuum between the lines, the fading of the emission line fluxes into the dissolved-level continuum at $\lambda < 3700$ \AA, and the bluest Balmer line (H15) are adequately reproduced in the RHD model superposition.  The wavelengths of the hydrogen series and a helium I line noted by \citet{HP91} are indicated.   }
       \label{fig:gflz}
       \end{centering}
   \end{figure}

We use the RH code \citep{Uitenbroek2001} with a 20-level hydrogen atom and the occupational probability modifications to the bound-bound and bound-free opacities that account for level dissolution at the Balmer limit \citep{Dappen1987, HM88, Nayfonov1999, Tremblay2009}.  The RH calculation setup is the same as described in \citet{Kowalski2017Broadening}.   These calculations are intensive because they involve a large numerical convolution at each atmospheric depth, and not every time-step in all models readily converges to a solution.  To demonstrate a representative solution with the two-component modeling approach from the previous section, we use atmospheric snapshots from the mF13-150-3 simulation at $t = 0.0, 0.4, 1.0, 2.0, 4.0, 6.0, 8.0, 9.8$~s and a snapshot from the m5F11-25-4 simulation at $t= 0.8$~s.  The F13 model spectra are coadded, and the preflare spectrum is subtracted from the two model components.  We then use the equation (Eq.\ \ref{eq:hgratio}),

\begin{equation} \label{eq:hgratio}
  \left( F^{\prime}_{\rm{H}\gamma} / \rm{C4170}^{\prime}\right)_{\rm{obs}} = \frac{X_{\rm{rel}} \times F^{\prime}_{\rm{H}\gamma \rm{,\ 5F11}} + F^{\prime}_{\rm{H}\gamma \rm{,\ F13}}}{X_{\rm{rel}} \times \rm{C4170}^{\prime}_{\rm{5F11}} + \rm{C4170}^{\prime}_{\rm{F13}}}
\end{equation}

\noindent to solve for $X_{\rm{rel}}$ given $(F^{\prime}_{\rm{H}\gamma} / \rm{C4170}^{\prime})_{\rm{obs}} = 20$.  The total two-component model flux spectrum is scaled to the continuum flare flux, C4170$^{\prime}$, in the observed spectrum.  For the two-component model combination above, a value of $X_{\rm{rel}} \approx 3.9$ is obtained, which is close to $4.5$ that is obtained from fitting the H$\gamma$ line profile (Table \ref{table:chisqr}).  We convolve the flare model with the spectral resolution of the data and show the result against the observations in Figure \ref{fig:gflz}, which demonstrates consistency with the observed Balmer jump flux ratio, the detailed merging of the line series wings, and in the bluest visible Balmer line.  Without the additional narrow-line flux from the 5F11, the highest balmer line in emission is H13, which is inconsistent with the observations.  Without the continuum and Balmer wing flux from the F13 model, the  Balmer jump ratio and the $F^{\prime}_{\rm{H}\gamma} / \rm{C4170}^{\prime}$ value from the 5F11 model are far larger (Table \ref{table:chisqr}) than measured from the observed spectrum.

  \subsection{Broadband Continuum Fitting} \label{sec:broadbandfits}
   In this section, we fit the observed continuum fluxes from the FUV to the red-optical during the early impulsive phase of the AD Leo Great Flare to compare to the results from fitting to the H$\gamma$ spectrum (Table \ref{table:chisqr}) and high-order series merging (Section \ref{sec:balmerlimit}).

\begin{figure}[ht!]
\begin{centering}
       \includegraphics[width=0.7\textwidth]{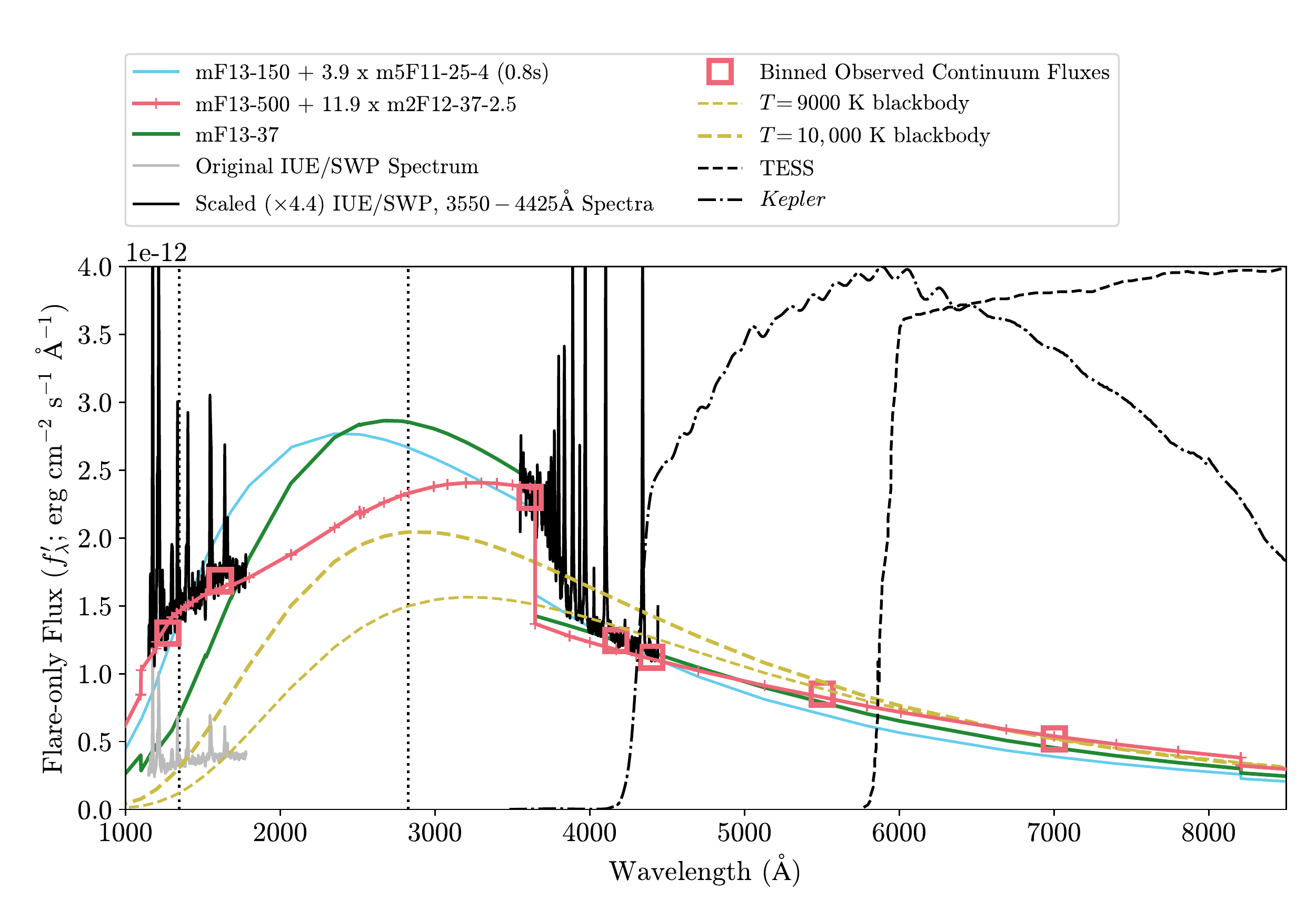}
       \includegraphics[width=0.37\textwidth]{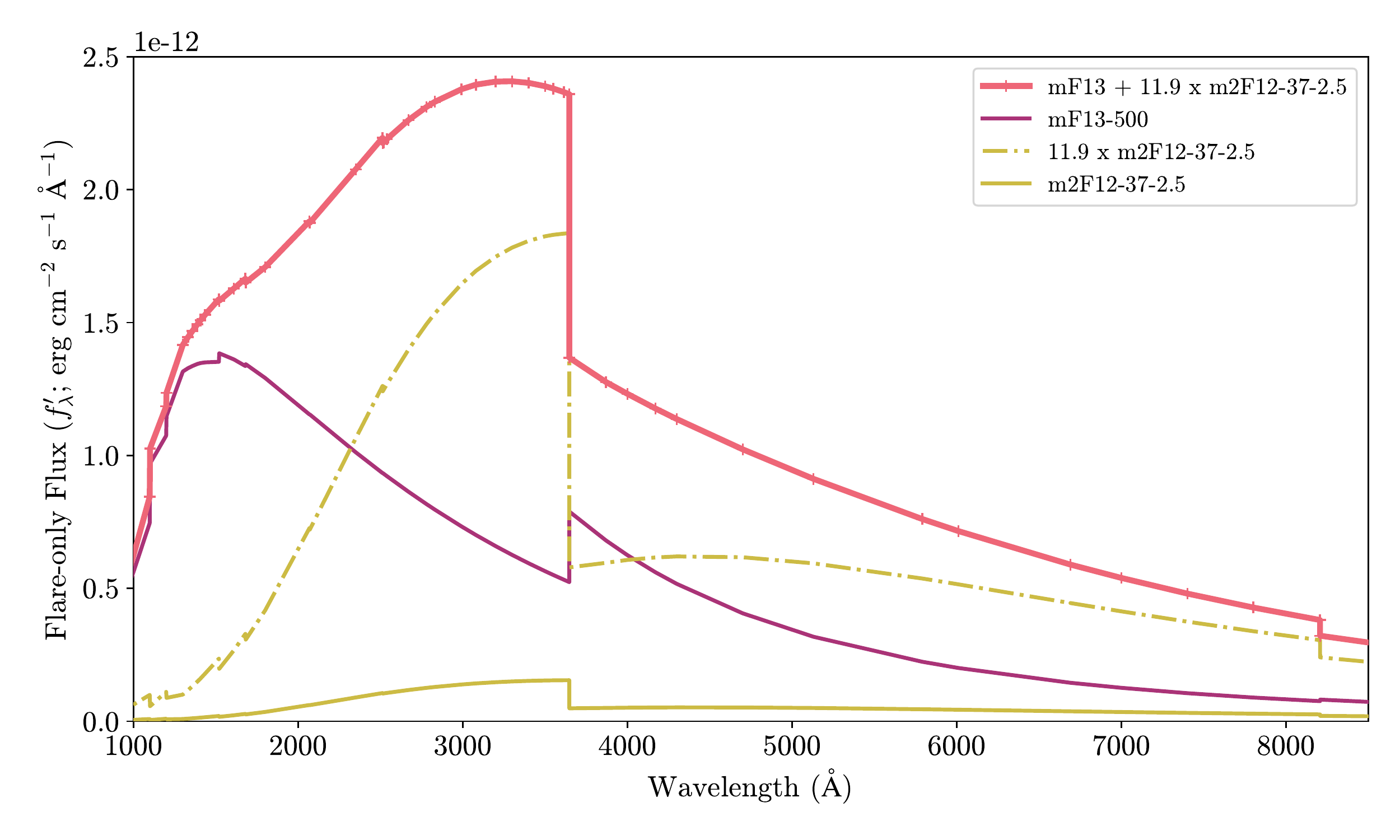}
       \includegraphics[width=0.3\textwidth]{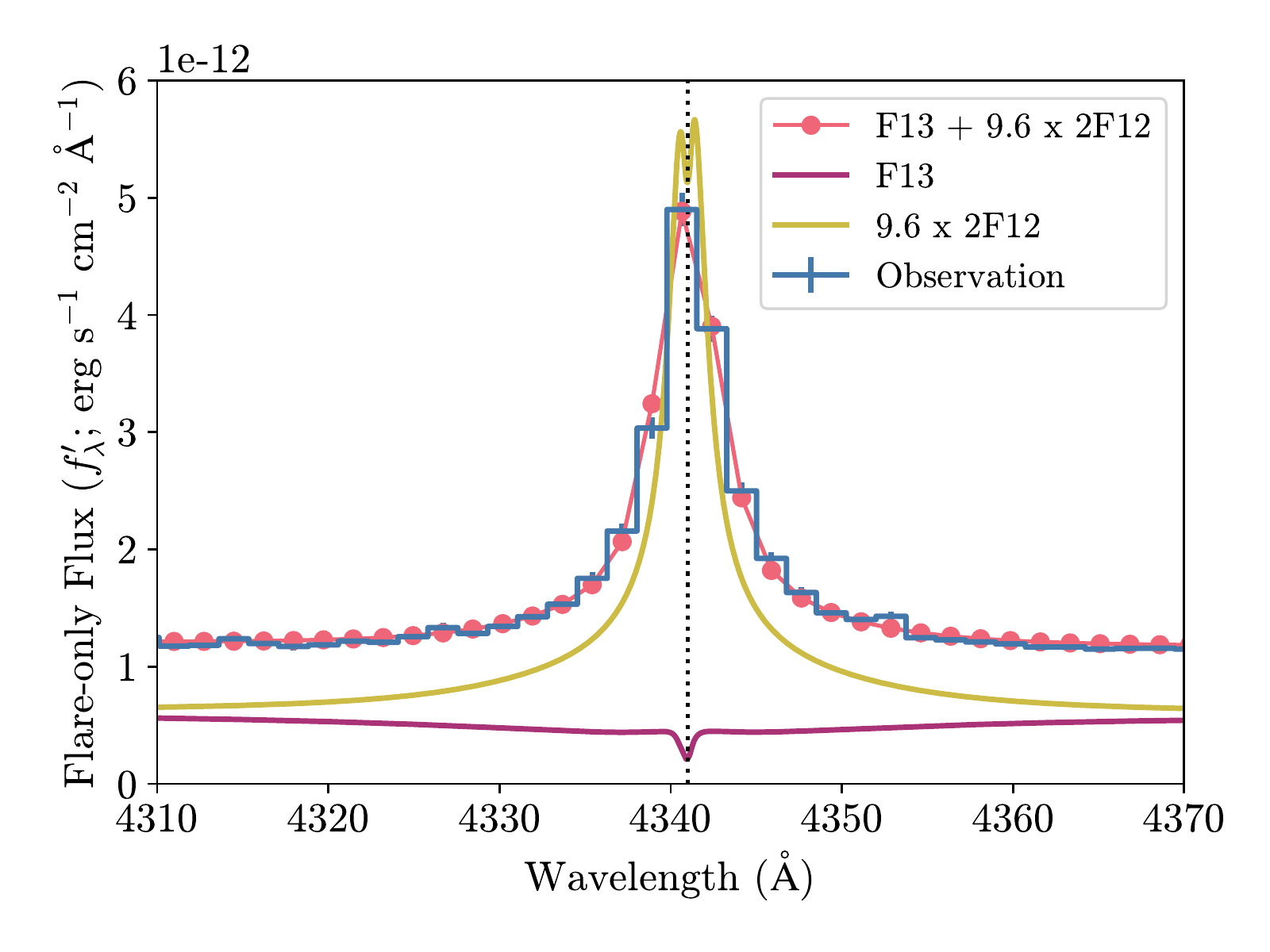}
       \caption{ (\textbf{Top}) Models of the early-impulsive phase broadband continuum flux distribution of the Great Flare of AD Leo compared to the IUE/SWP spectrum, optical ground-based spectrum, and broadband $V$ and $R$ photometry.  All observed fluxes have been adjusted to the synthetic $U$-band flux according to the broadband distribution at $t=0-900$~s in Fig.\ 11 of \citet{HF92}.  The wavelength-binned, flare-only fluxes that are used to fit the models are indicated by square symbols with a best-fit, two-component RHD model continuum spectrum shown as the solid red line.  Other model predictions are scaled to the observations as follows:  the blackbody functions are scaled to the $R$ band flux observation, and the other two RHD models are scaled to the average continuum flux at $\lambda = 4155-4185$\AA. (\textbf{Bottom, left}) Individual model components in the best-fit mF13-500-3 $+ 11.9 \times $ m2F12-37-2.5 combination, which compares the relative contributions to the spectral luminosity of the continuum radiation in the Great Flare.  The m2F12-37-2.5 model component is also shown without scaling by the best-fit filling factor to facilitate direct comparison to the radiative surface flux of the mF13-500-3.  (\textbf{Bottom, right})  Best-fit H$\gamma$ line profile model using the mF13-500-3 and m2F12-37-2.5 flux spectra gives a similar value of $X_{\rm{rel}}$ as for the fits to the broadband continuum fluxes.   }
       \label{fig:nuvgf}
       \end{centering}
   \end{figure}

 Figure \ref{fig:nuvgf} shows a comparison of the detailed continuum fluxes for several combinations of the models that satisfactorily explain the early-impulsive phase, blue-optical spectrum of the Great Flare of AD Leo.   
 A representative RHD model combination (mF13-150-3, m5F11-25-4 ($t=0.8$s)) from Section \ref{sec:balmerlimit} exhibits a peak at $\lambda \approx 2350-2400$ \AA\ followed by a turnover toward shorter wavelengths.   Qualitatively, these properties are consistent with FUV constraints of this superflare and other, smaller flares from AD Leo \citep{Hawley2003}, but the IUE/SWP observation allows a more detailed comparison. 
To adjust the flux calibration of the data for the different exposure times between the early-impulsive phase, IUE/SWP spectrum ($t < 900$~s in Fig.\ \ref{fig:gflc}) and the $\lambda > 3560$ \AA\ ground-based spectrum ($t = 542 \pm 90$~s in Fig.\ \ref{fig:gflc}; see Section \ref{sec:observations}), we apply the relative scaling between the $U$-band and the FUV continuum flux at $\lambda \approx 1600$ \AA\ within the first 900~s of the flare that is presented in the upper left panel of Fig.\ 11 of \citet{HF92}.  The relative scale factor ($1.3$) is used to adjust the lower envelope of the SWP continuum flux relative to a synthetic $U$-band flux that we calculate from the blue-optical spectrum.  The scaled and original IUE/SWP spectra are shown in Figure \ref{fig:nuvgf}.  We also include the $V$- and $R$-band photometry from the same figure in \citet{HF92} and apply the scaling in the same way as for the FUV continuum.  We calculate two-parameter, linear least-squares fits to the seven flare-only flux measurements of the continuum in the Great Flare\footnote{Instead of $U$ and $B$-band photometry used for model fitting in \citet{HF92}, we use C3615$^{\prime}$, C4170$^{\prime}$, and C4400$^{\prime}$ calculated from averages of the continuum fluxes at $\Delta \lambda \approx 30$ \AA\ around $\lambda = 3615, 4170, 4400$\AA, respectively.}.  Minimizing $\chi^2$ (Section \ref{sec:fits}) gives several combinations of models with very large low-energy cutoffs ($E_c = 350 - 500$) as the $X_{\rm{F13}}$ model component superimposed with the m2F12-37-2.5 model spectrum.  Note, the m2F12-37-2.5 model is the only simulation in our \texttt{RADYN} grid with such a hard,  $\delta < 3$, electron beam power-law index.

The best-fit superposition of the mF13-500-3 and m2F12-37-2.5 radiative flux spectra is shown in Figure \ref{fig:nuvgf} (top panel) with $X_{\rm{rel}} \approx 11.9$.  Notably, this fit comprehensively accounts for the slope of the lower-envelope of the FUV flare spectrum, the Balmer jump strength, and the optical continuum constraints.  The middle, left panel of Figure \ref{fig:nuvgf} shows the contributions of the individual model components to the spectral luminosity of the flare continuum from the top panel.  The mF13-500-3 accounts for most of the FUV continuum luminosity, whereas the Balmer jump in the m2F12-37-2.5 contributes a larger fraction in the NUV and in the $U$ band.  At optical wavelengths,  relative contributions to the blue continuum spectral luminosity are about equal, but the lower beam-flux model is larger toward near-infrared wavelengths.  The comparisons of the surface flux spectra without adjustments by the best-fit filling factors emphasize that the F13 model is the much brighter source at all wavelengths.   A fit using these two model component to the observed H$\gamma$ line profile (Section \ref{sec:fits}) is shown in Figure \ref{fig:nuvgf} (bottom, right panel);  the fit is excellent and, moreover, returns a similar, independent estimate for the parameter $X_{\rm{rel}} = 9.6$ (Table \ref{table:chisqr}).

The fully-relativistic electron beam parameters of the mF13-500-3 beam are rather extreme, but they are not without precedent and sufficient semi-empirical necessity. \citet{Kowalski2017Broadening} used a superposition of three \texttt{RADYN} simulations to 
model the decay phase spectra of a superflare from the dM4.5e star YZ CMi.  A \texttt{RADYN} model with a constant electron beam energy flux injection of $2\times 10^{12}$ erg cm$^{-2}$ s$^{-1}$, a low-energy cutoff of $E_c =500$ keV, and a power-law index of $\delta =7$ was used to explain the spectra of a secondary flare event, which exhibit features that are similar to an A-type star photospheric spectrum:  namely, broad Balmer lines and a Balmer jump ``in absorption'' \citep[see also][]{Kowalski2012SoPh, Kowalski2013}.  Secondary flare events in the decay phase of a large flare from the young G-dwarf, EK Dra, were reported in \citet{Ayres2015} to exhibit a response in only the FUV continuum.  Finally, we note that increasing the value of $X_{\rm{rel}}$ after the peak flare phase may be able to explain the relatively rapid nature of the FUV continuum radiation that has been reported in other M dwarf flares \citep{Hawley2003, MacGregor2020}.

\section{Discussion} \label{sec:discussion}

\subsection{Summary of Fitting Results} \label{sec:summary}

We fit the Great Flare impulsive phase (rise/peak) spectrum using simulations of electron beam heating from a new grid of \texttt{RADYN} flare models.  The data require two, independent RHD model components, resulting in relative filling factors ($X_{\rm{rel}}$) of the components between $\approx 1 - 10$, with the lower beam flux model component exhibiting the larger filling factor.  The time-evolution of the simulated atmosphere and emergent radiative flux spectra over each heating pulse is included in these comparisons to the data. 
Several examples of fits were presented and discussed.   The shape of the H$\gamma$ line profile far into the wings and nearby continuum flux
constrains combinations of a high-flux model (F13) with large low-energy cutoff values ($E_c = 85 - 500$ keV) and a lower-flux model (5F11--2F12) with smaller low-energy cutoff values ($E_c = 25-37$ keV).    In the (mF13-85-3, m5F11-25-4) spectral luminosity, most of the  H$\gamma$ wing broadening and blue-optical continuum radiation is attributed to the F13 model  component. In the (mF13-200-3, mF12) model fits, most of the blue-optical continuum luminosity  is due to the F13 spectrum, but much of the wing broadening can be attributed to the lower beam flux, F12, component.  This is qualitatively consistent with the modeling results from \citet{Namekata2020}, who found that similar F12 electron beam models produce satisfactory agreement in the broadening of the H$\alpha$ line in a superflare.  In their work, however, detailed comparisons to the spectra of the blue-optical continuum radiation were not possible.  

We examined the prediction of one of the fits to the Great Flare H$\gamma$ line against the spectrum of the hydrogen series at the Balmer limit; there is remarkable agreement with the highest-order Balmer line in emission and with other features in the rise/peak spectrum (Figure \ref{fig:gflz}).  Models are also independently fit to the broadband photometry and spectral distribution during the first 900s of the Great Flare, and the superposition of spectra from the mF13-500-3 (or mF13-350-3) and the m2F12-37-2.5 RHD model components gives an excellent fit;  moreover, this fit results in about the same relative filling factor as inferred from the H$\gamma$ line profile fitting.  For this combination of models, the relative contributions to the optical continuum luminosity are comparable, but the $E_c=500$ keV model dominates the FUV flare luminosity.  In all model combinations, the F13 model component produces the brightest continuum surface flux.

In this section, we use the results from the fits to discuss the implications for models of the NUV radiation environment of the habitable zones of low-mass flare stars (Section \ref{sec:hz}).   Then, we examine a high-spatial resolution image of a widely-studied X-class solar flare to speculate on the origin of these two spectral components in terms of solar flare phenomenology (Section \ref{sec:interpretation}).  We show how the relative filling factors of the two model components are consistent with the relative areas of solar flare kernels and ribbon intensities, respectively, in the impulsive phase of this solar flare.  In Section \ref{sec:futurework}, we discuss further empirical investigation to anchor the two-component continuum and H$\gamma$ broadening models of stellar flares in reality.

\subsection{The NUV Radiation Field in Habitable Zones of Low-Mass, Flare Stars} \label{sec:hz}

The detailed RHD models provide insight into the magnitude of the possible systematic errors for the inferred NUV radiation field in the habitable zones of low-mass flare stars. 
The RHD spectra in the NUV reveal that simple extrapolations from flare photometry in the red-optical and near-infrared (e.g., from the \emph{Kepler} or TESS bands) that do not account for the Balmer jump strength, may result in rather large systematic modeling errors.  We scale a $T=9000$ K and $T=10,000$ K blackbody to the observed $R$-band flux of the Great Flare in Figure \ref{fig:nuvgf} (top panel). Compared to the RHD models, the blackbody models under-predict the $\lambda = 1800-3646$ \AA\ flare-only flux by factors ranging from $1.2$ to $2.0$.   The peaks and slopes of the UV and $U$-band continuum spectra are largely in disagreement as well. Scaling all models to a common RHD continuum flux at a redder continuum wavelength, $\lambda = 7810$ \AA, that is closer to the central wavelength of the TESS white-light band (Figure \ref{fig:nuvgf}) generally results in larger underestimates of the NUV continuum flare-only flux by factors up to $2.6$.  The inadequacies of single-temperature, blackbody models are even more evident at $\lambda = 1100- 1800$ \AA\ and in the expected amount of Lyman continuum fluxes at $\lambda \lesssim 911$ \AA\ (not shown) that are present in the RHD model spectra.

The recent laboratory experiments of \citet{Abrevaya2020} measured survival curves of microorganisms that were  irradiated by sustained fluxes of monochromatic NUV light at $\lambda = 2540$ \AA.  In the worst-case scenario of direct irradiation, they found that a large UV-C flux from a superflare in the habitable zone \citep[$d=0.0485$ au;][]{Anglada2016} of the dM5.5e star Proxima Centauri fails to terminate biological function in a small but non-negligible fraction of the initial sample.  The UV-C flux\footnote{For continuity with these studies, we momentarily express quantities in S.I.\ units.} of $92$ W m$^{-2}$ was calculated by scaling a $T=9000$ K blackbody curve to the peak magnitude change in the Evryscope $g^{\prime}$ bandpass as described in \citet{Howard2018};  we refer the reader to \citet{Law2015} and \citet{Howard2019} for details about the Evryscope survey.  We estimate that the peak $B$-band ($\lambda = 3910 - 4890$ \AA) luminosity of the Great Flare of AD Leo was at least a factor of three larger than the $g^{\prime}$-band ($\lambda \approx 4050 - 5500$ \AA) peak luminosity of the Proxima Centauri superflare.  If a flare as luminous as the Great Flare (and the same in all other regards) were to occur on Proxima Centauri, the RHD models in Figure \ref{fig:nuvgf} predict  UV-C, impulsive-phase, habitable-zone fluxes of $800-1000$ W m$^{-2}$.  This range is rather similar to the habitable-zone, UV-C fluxes inferred in \citet{Howard2020} using extrapolations from much higher temperature blackbody fits to broadband optical and near-IR photometry.  As \citet{Howard2020} discuss, it would be interesting for laboratory experiments to determine whether there is an upper limit to the UV-C flux at which a  microbial population achieves a steady-state survival fraction.

The pioneering study of \citet{Segura2010} combined the multi-wavelength AD Leo flare spectra for empirically-driven photochemistry and surface UV dosage modeling of an Earth-like planet in the habitable zone at $d=0.16$ au from a dM3 star. In their approach, \citet{Segura2010} used the first IUE/LWP (NUV) spectrum available (starting at 5:00 UT in Figure \ref{fig:gflc}) to bridge the blue-optical and IUE/SWP spectra during the early impulsive phase in the first 900~s of Figure \ref{fig:gflc}.  This approach assumes that the peak impulsive-phase NUV flare spectrum is the same as that at the end of the fast decay and start of the gradual decay phase in this event.  At longer wavelengths, this assumption is not consistent with analyses of more recent, time-resolved spectra \citep{Kowalski2013}.  However, we think that this approach is reasonably justified given the vagaries inherent in such spectral observations with relatively long exposure times that are not contemporaneous within the Great Flare.

 Our scaling of the IUE/SWP spectrum (Figure \ref{fig:nuvgf}) follows a different approach and is consistent with the relative surface fluxes at $\lambda \approx 1600$ \AA\ and the $U$ band that are shown in the upper left panel of Fig.\ 11 of \citet{HF92} and in Table 5 of \citet{HP91}.   In the IUE/LWP decay phase spectra of the Great Flare, the ratio of the $\lambda = 2800$ \AA\ to $\lambda = 2000$ \AA\ continuum fluxes is $\approx 5$ \citep[][see also \citet{Segura2010}]{HP91}, which effectively force the FUV continuum flux to a lower value relative to the fluxes at longer wavelengths in the NUV and $U$ band.   In our best-fit (mF13-500-3, m2F12-37-2.5) continuum flux model of the early impulsive phase (Figure \ref{fig:nuvgf}), the $\lambda = 2800$ \AA\ to $2000$ \AA\ continuum flux ratio is only $1.3$.  Thus, one expects the wavelength-integrated, UV-C flux of this model combination to be a factor of $\approx 1.5$ larger than the empirical model of \citet{Segura2010}, assuming equal top-of-the-atmosphere fluxes at $\lambda \approx 2800$\AA.  However, further comparison reveals that the \citet{Segura2010} composite flare spectrum is similarly flat at $\lambda \gtrsim 2400$\AA.  This effect is apparently due to the large number of blended (and saturated) Fe II and Mg II emission lines, which are generally much more prominent relative to the flare continuum radiation in the decay phase \citep{Kowalski2019HST}.  This coincidence is fortuitous for many follow-up studies \citep[e.g.,][]{Venot2016, Tilley2019} that have adopted the composite NUV flare spectra from \citet{Segura2010} for photochemistry modeling: the pseudo-continuum of blended, saturated lines in the decay phase of the Great Flare mimics the shape of our RHD model continuum distribution that best reproduces the available observations in the rise and peak phases.

  \subsection{A Solar Flare ``Kernel$+$Ribbon'' Interpretation of the Great Flare Rise/Peak Phase} \label{sec:interpretation}

\begin{figure}[ht!]
   \begin{centering}
       \includegraphics[width=1.0\textwidth]{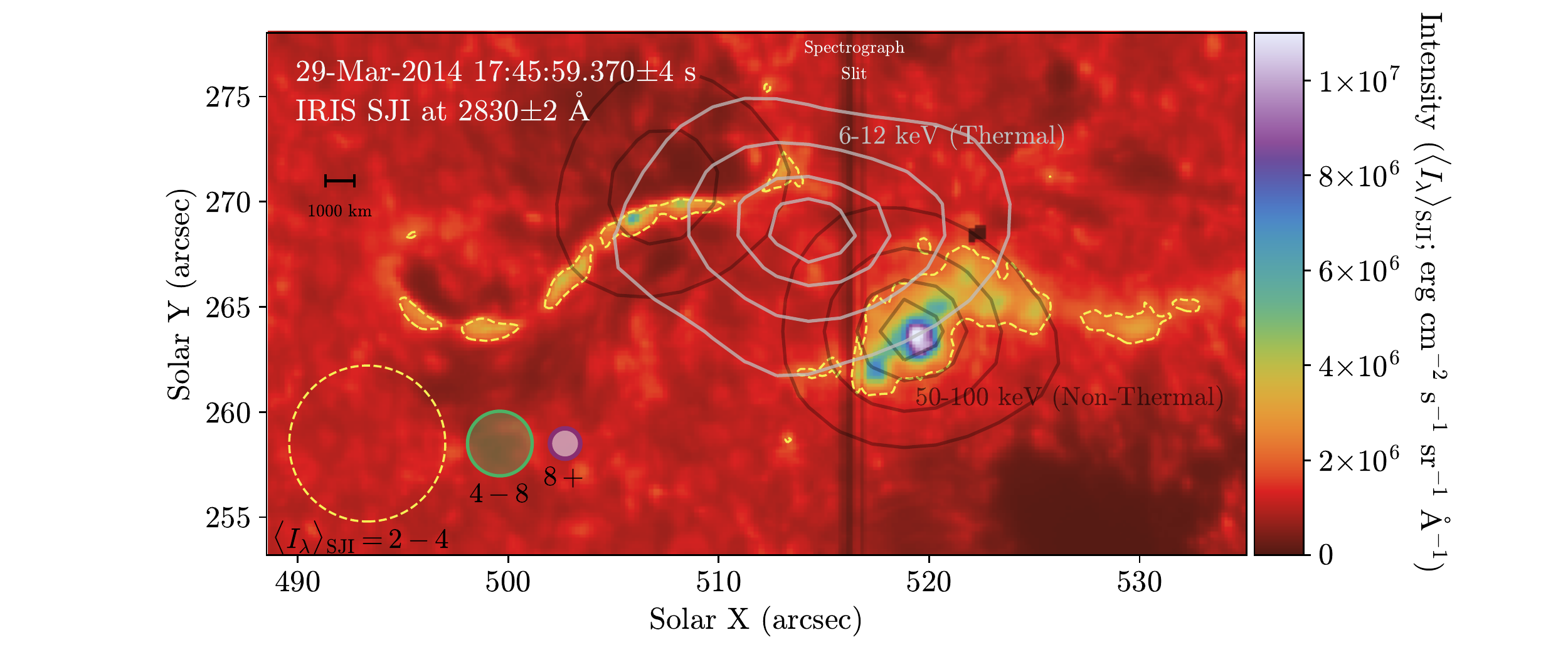}
   \caption{ IRIS SJI 2832 image during the hard X-ray impulsive phase of the 2014 Mar 29 solar flare.  The spatial resolution of the IRIS image is $0.\arcsec 4$ ($0.\arcsec 167$ pix$^{-1}$; $724$ km arcsec$^{-1}$).   The projected, exclusive areas of $2.3 \times 10^{17}$, $3.9 \times 10^{16}$, and $8.6 \times 10^{15}$  cm$^2$ correspond to the intensity ranges indicated in the figure below the equivalent circular areas.    Note, an excess image formed by subtracting the image from 150~s earlier reveals much fainter emission;  in this case, an excess threshold value of $6 \times 10^5$ \ilam\ \citep{Kowalski2017Mar29} show that the faintest parts of the ribbons extend over an area of $\approx 5\times 10^{17}$ cm$^2$.     The RHESSI X-ray contours are plotted at 25, 50, 75, and 90\% of the maxima.       }
   \label{fig:iris}
   \end{centering}
\end{figure}

In this section, we argue that the results from the two-component model RHD fits are ostensibly consistent with the relative areas of high-intensity and medium-intensity sources in the impulsive phase of well-studied solar flare.
The IRIS SJI 2832 image during the impulsive phase of the 2014 March 29 X1 solar flare is shown in Figure \ref{fig:iris}.    We calculate the areas corresponding to several intensity ranges in the SJI 2832 image:  $\langle I_{\lambda} \rangle_{\rm{SJI}} = 2-4 \times 10^6$ \ilam\ (faint threshold),  $4-8 \times 10^6$ \ilam\  (medium threshold), and  $\ge 8\times10^6$ \ilam\ (bright threshold).   The bright threshold selects the pixels corresponding to the bright kernel \citep[which is referred to as BK\#1 in ][]{Kowalski2017Mar29}, and the medium threshold corresponds to the elongated ribbons on both sides of the brightest kernel\footnote{The medium threshold approximately corresponds to the ``high thresh'' area calculated from the excess intensity images in \citet{Kowalski2017Mar29}.}. In Figure \ref{fig:iris}, the dashed contour outlines the faint threshold area, the green-colored pixels correspond to the medium threshold pixels, and the purple and white pixels isolate the bright threshold kernel. We sum the exclusive areas within these three intensity ranges, and the equivalent circular areas that correspond to the assumed stellar footpoint geometry (Eq. \ref{eq:superpos}) are illustrated in the bottom left of the figure.  The ratios of these areas are $\approx$ 25:5:1 for the faint:medium:bright areas, respectively.  Coincidentally, the ratio of medium:bright areas is 5:1, and the ratio of faint:medium brightness areas  is also 5:1.   These ratios are very similar to the areal ratios ($X_{\rm{rel}}$; Table \ref{table:chisqr}) that we inferred between the higher flux and lower flux models through our spectral fitting to the AD Leo Great Flare.   Thus, we attribute the  two model components as representing a bright kernel -- or several bright kernels -- and fainter ribbons over a larger area.

The two RHD model components could also represent the faint-intensity and medium-intensity areas, respectively, which exhibit an areal ratio of 5:1.  To justify this interpretation as the less plausible analogy for stellar flares, we bring in analyses of a solar, \texttt{RADYN} flare model (``c15s-5F11-25-4.2'') from \citet{Kowalski2017Mar29} and \citet{Kowalski2022}. \citet{Kowalski2017Mar29}  synthesized the SJI 2832 intensity from this model, accounting for the emission lines and continuum response in this bandpass.  At the brightest times of the 5F11 model ($t \approx 4$~s), they calculate a synthetic SJI 2832 intensity of $\langle I_{\lambda} \rangle_{\rm{SJI}} \approx 10^7$ \ilam, which is consistent with the brightest pixels in Figure \ref{fig:iris}.  Since this study, a detailed identification of the emission lines that contribute to the SJI 2832 data in solar flares has been presented \citep[][see also \citealt{Kleint2017}]{Kowalski2019IRIS}, and several updates to the atomic physics of Fe II have been implemented (which are to be described elsewhere in detail).  The new calculations result in fainter emission line flux but similar redshift evolution of the Fe II lines.  Averaging the solar c15s-5F11-25-4.2 model from \citet{Kowalski2022} over a simulated SJI 2832 exposure time of 8~s, so as to be directly comparable to the brightest pixels in the data in Figure \ref{fig:iris}, results in a synthetic model intensity of only $\langle I_{\lambda} \rangle_{\rm{SJI}} \approx 5 \times 10^6$ \ilam.   This intensity is above the medium-intensity threshold area corresponding to the green-colored ribbon pixels in Figure \ref{fig:iris}, but it is  not nearly as bright as the most intense pixels.  The IRIS raster clearly ``steps over'' the brightest kernel\footnote{We further confirm this by inspecting the Mg II slit jaw images:  though large regions of the ribbon are saturated, most of the saturation occurs away from the slit.} in Figure \ref{fig:iris} \citep[see Fig.\ 1 in \citet{Kowalski2017Mar29} and the discussion in][]{Kleint2016}.   Thus, the 5F11 electron-beam, solar flare modeling with a small, low-energy cutoff is apparently most appropriate for the medium-intensity ribbons instead of the fainter ribbons that extend beyond the hard X-ray contours or the brightest pixels at the centroid of the hard X-ray contours.  The faintest ribbon intensity in the impulsive phase\footnote{The next SJI 2832 image corresponds to the beginning of the fast decay phase of the hard X-rays, and the faint intensity threshold clearly selects a large area in the ``wakes'' of the bright ribbons.  In these wakes, the emission lines may exhibit broad, nearly symmetric profiles as the red-wing asymmetries have coalesced with the line component near the rest wavelength \citep{Graham2020} while the flare continuum intensity is still at a detectable level in the IRIS NUV spectra \citep{Kowalski2017Mar29, Panos2018, Zhu2019}.}   may correspond to locations of impulsive energy deposition by thermal conduction \citep{Battaglia2015, Ashfield2022} and/or XEUV backheating over a large area \citep{Fisher2012}.  The latter has been investigated in detail with 1D non-LTE models for the Great Flare data \citep{HF92}; we speculate that radiative backheating from an arcade of hot loops \citep[e.g.][]{Kerr2020} may account for additional Ca II K line flux in the model spectra in Figure \ref{fig:gflz} (top panel).

Is there evidence that a much stronger source of heating than a 5F11 beam contributes to the brightest SJI 2832 kernel pixels in this solar flare? 
Using an even brighter intensity threshold of $\langle I_{\lambda} \rangle_{\rm{SJI}} = 10^7$ \ilam\ to mask the solar flare kernel in Figure \ref{fig:iris} gives an area of $3.3 \times 10^{15}$ cm$^{-2}$, or four IRIS pixels.  Dividing this area into the nonthermal electron power ($8 \times 10^{27}$ erg s$^{-1}$) above $20$ keV  that is inferred through standard collisional thick target modeling in \citet{Kleint2016} gives an injected electron beam flux of $2 \times 10^{12}$ erg cm$^{-2}$ s$^{-1}$ (2F12).  This is not as high as the maximum injected beam fluxes in the F13 models, but it is much larger than typically considered in solar flare RHD modeling.  This line of reasoning implies that a much higher beam flux model is a more appropriate collisional thick target inference for the brightest SJI 2832 kernel in this solar flare.  For these large beam fluxes, however, the standard assumptions in collisional thick target modeling of the hard X-ray footpoints are not applicable when the ambient coronal densities are small \citep{Krucker2011}.
Although the RHESSI sources are largely unresolved \citep[with a spatial resolution of $3.\arcsec 6$, or $5.4 \times 10^{16}$ cm$^{2}$ at the Sun;][]{Battaglia2015}, the spatially integrated hard X-ray and IRIS SJI 2832 powers provide upper limits on necessitated modifications to the thick-target physics \citep[e.g.,][]{Kontar2008, Brown2009, Kontar2012, Hannah2013, Alaoui2017, Allred2020} that are implemented in future modeling of the heterogeneous atmospheric response within the hard X-ray source contours.  \citet{Graham2020} investigated the deficiencies in chromospheric condensation model predictions of the red-wing asymmetry evolution of Fe II flare lines in IRIS spectra.  Resolving the disagreements, and drawing on implications for the standard collisional thick target inferences of beam parameters, would greatly enhance the realism of the analogous stellar flare RHD component with small low-energy cutoff values.

    \subsection{Future Observational Constraints} \label{sec:futurework}
\citet{Graham2020} used two intensity thresholds in faster-cadence SJI 2796 imaging of a different X1 solar flare to quantify newly brightened areas as a function of time.  The ratios of these areas are $\approx 10:1$, and the areal evolution is rather similar to the timing of the spatially integrated, hard X-ray emission peaks from Fermi/GBM.  Further
verification of the heterogeneity between kernel and the medium-brightness ribbon pixels are clearly needed from solar flare spectral observations.  One such unexplored constraint is the NUV and FUV continuum evolution from IRIS flare spectra.  In Figure \ref{fig:nuvgf} (top), we show the locations of two continuum windows in the IRIS spectra around $\lambda \approx 2826$ \AA\ and $\lambda \approx 1349$ \AA. The two spectral components obtained  from our fits exhibit distinct, time-dependent C2826$^{\prime}$/C1349$^{\prime}$ emergent intensity ratios.  A detailed investigation of the relative continuum intensities for a large sample of  solar flares would help to determine the heterogeneity of the atmospheric response between the brightest kernels and nearby bright ribbons.

Reality checks could also be attained through spatially resolved characterization of the hydrogen Balmer line broadening along the slit length of observations of solar flares with the Visible Spectropolarimeter \citep[ViSP;][]{deWijn2022} on the Daniel K.\ Inouye Solar Telescope \citep[DKIST;][]{Rimmele2020}.  Our stellar flare phenomenological modeling paradigm (Section \ref{sec:interpretation}) implies that the continuum-subtracted effective widths of the H$\gamma$ emission line \citep{Kowalski2022} from the emergent intensity spectra of the brightest kernels should exhibit much larger effective widths (Table \ref{table:models}, rightmost column) than the medium-brightness ribbon component.  In solar observations, the pixels with the largest H$\gamma$ effective widths should also show the brightest blue-optical continuum intensity. 
A statistical classification of hydrogen line spectra should reveal distinct components that correlate with timing and position along the solar flare ribbons, similar to the groupings that were reported for a large sample of Mg II flare lines in IRIS spectra \citep{Panos2018}.

On the stellar side, high-cadence spectral observations of low-mass stellar flares at $\lambda = 1800-3200$ \AA\ during the impulsive phase would clarify how the NUV continuum flux peaks and turns over into the FUV in events like the Great Flare, which exhibits a small Balmer jump and  a highly-impulsive, broadband temporal evolution.  The Cosmic Origins Spectrograph on the \emph{Hubble Space Telescope} provides such an opportunity:  the G230L grating with a central wavelength at $\lambda =3000$ \AA\ gives simultaneous spectral coverage at $\lambda = 1700-2100$ \AA\ and $\lambda = 2800-3200$ \AA, which would provide the necessary observations to test the RHD models.  Recently, \citet{Kowalski2019HST} reported on flare spectra from the Cosmic Origins Spectrograph (using a different central wavelength) and constrained the peak continuum flux to the $U$ band.  It was argued that  these events are gradual-flare (GF) events with large Balmer jumps and large line-to-continuum ratios, which are in stark contrast to the measured quantities from the Great Flare optical spectra.  We briefly comment that our two-component modeling can readily reproduce the properties of these gradual-type flare events as well.  For example, a two-component model consisting of the mF13-150-3 and the m5F11-25-4 spectra with $X_{\rm{rel}} \approx 90$ results in a Balmer jump ratio of $3.9$ and a value of $F^{\prime}_{\rm{H}\gamma} /\rm{C4170}^{\prime} \approx 150$, which are consistent with the quantities from HST-1 in \citet{Kowalski2019HST}.  A parameter study of the detailed hydrogen line broadening and NUV spectra are planned as the subject of Paper II in that series.

\section{Summary \& Conclusions}  \label{sec:conclusions}

We have comprehensively modeled the multi-wavelength spectra during the rise and peak phase of the Great Flare of AD Leo \citep{HP91}.  We have shown that fitting two RHD spectral components to the detailed properties of the hydrogen series using an updated treatment of the pressure broadening, combined with a mechanism that heats deep chromospheric heights to $T \gtrsim 10^4$ K, is readily feasible with satisfactory statistical significance.   This semi-empirical modeling approach accounts for the evolution of height- and wavelength-dependent emission line and continuum opacities in the flare atmosphere, which is self-consistently calculated in response to high-flux electron beam heating.  A simulation \citep{Kowalski2015, Kowalski2016} with a large electron beam flux  and the largest  low-energy cutoff value range ($\lesssim 40$ keV) that is inferred from solar flare data \citep{Holman2003, Ireland2013} produces a dense chromospheric condensation and hydrogen Balmer wings that are far too broad compared to the observation.  Models that exhibit a large ($\gtrsim 85$ keV), low-energy cutoff and high electron beam flux ($\approx 10^{13}$ erg cm$^{-2}$ s$^{-1}$) are able to explain the observed continuum distribution and highly broadened Balmer line wings that are within the constraints of the Balmer H$\gamma$--H16 emission line series. Large, low-energy cutoffs are sometimes inferred in the so-called ``late impulsive peaks'' in solar flares \citep{Holman2003, Warmuth2009}, and significant progress has been made to improve the hard X-ray modeling of these events beyond the physics in the standard collisional thick target model \citep{Alaoui2017}. 

A second, lower electron beam flux model exhibiting more similarities to nonthermal electron parameters that are typically inferred in collisional thick target modeling of hard X-ray data of solar flares (e.g., fluxes of $5 \times 10^{11}$ erg cm$^{-2}$ s$^{-1}$, low-energy cutoffs of $E_c \approx 25$ keV) is necessary to fit the narrower hydrogen Balmer emission peak fluxes and account for the bluest Balmer line in the AD Leo Great Flare spectrum.  We suggest that this second component represents heterogeneity of nonthermal beam injection and the differences between bright, larger area ribbons and brightest kernel morphologies that are readily seen in the impulsive phase of solar flare imagery.   However, further verification is needed from solar observations: specifically, comparisons of hydrogen spectra at different locations in early flare development are critical.  The implementation of this two-component, semi-empirical RHD modeling approach to Balmer line profiles of other M dwarf flares with higher resolving-power, echelle observations is underway (Kowalski et al.\ 2022, in preparation; Notsu et al.\ 2022, in preparation) and will further constrain plausible linear combinations of RHD model spectra.

The effects of transient UV radiation during flares is a relatively new topic in the study of exoplanet habitability (e.g., surface dosages) and atmospheric photochemistry (e.g., ozone photodissociation).  These studies would benefit from new NUV spectral observations of stellar flares.
The semi-empirical combination of RHD model spectra that are fit to the Great Flare observations predict unexpected properties of the NUV continuum spectra of impulsive-type M dwarf flares with small Balmer jumps, highly broadened Balmer lines, and small line-to-continuum ratios.  We conclude that small Balmer jumps, which appear as relatively small deviations from a $T = 9000$ K or $T = 10,000$ K blackbody fit to optical data in some flares, actually are consistent with much more energetic sources of ultraviolet radiation than previously thought were possible from solar and stellar chromospheres.

\section*{Conflict of Interest Statement}

The authors declare that the research was conducted in the absence of any commercial or financial relationships that could be construed as a potential conflict of interest.

\section*{Author Contributions}

AFK contributed by analyzing the AD Leo flare data, running the RHD models, and performing the least squares fitting analysis.  AFK lead the analysis of the solar flare imaging, and he wrote the manuscript and produced the figures.

\section*{Funding}
AFK acknowledges funding support from NSF Award 1916511, NASA ADAP 80NSSC21K0632, and NASA grant 20-ECIP20\_2-0033.

\section*{Acknowledgments}
We thank two anonymous referees for their detailed comments that helped improve this work.
AFK thanks Suzanne L. Hawley for many stimulating conversations about flare spectra, flare modeling, and the AD Leo Great Flare data.  AFK thanks Joel C. Allred and Mats Carlsson for access to and assistance with the \texttt{RADYN} code, and Han Uitenbroek for access to and assistance with the \texttt{RH} code.  AFK thanks Pier-Emmanuel Tremblay for access to the TB09$+$HM88 hydrogen broadening profiles.  Many discussions with Rachel A. Osten, Clara Brasseur, Isaiah Tristan, Ward Howard, and John P. Wisniewski about NUV and optical flare continuum radiation were important for the motivation and development of the ideas in this work.  AFK thanks Eduard Kontar for helpful discussions about the collisional thick target model of solar flare hard X-ray emission.  AFK thanks Eric Agol for suggesting ten years ago to explore large, low-energy cutoffs in flare heating models.  Many discussions about Balmer jump solar flare spectra and instrumentation with Gianna Cauzzi, Hoasheng Lin, and Tetsu Anan led to refining the ideas about the solar-stellar connection presented in this work.
IRIS is a NASA small explorer mission developed and operated by LMSAL with mission operations executed at NASA Ames Research Center and major contributions to downlink communications funded by ESA and the Norwegian Space Centre.

\bibliographystyle{Frontiers-Harvard}

\bibliography{frontiers_v5}

\end{document}